# Global Uncertainty-Sensitivity Analysis on Mechanistic Kinetic Models: From Model Assessment to Theory-Driven Design of Nanoparticles

M. Reza. Andalibi[*,†,‡], Paul Bowen[‡], Agnese Carino[†], Andrea Testino[†]

[†] Paul Scherrer Institute (PSI), ENE-LBK-CPM, Villigen-PSI, Switzerland.

[‡] École polytechnique fédérale de Lausanne (EPFL), STI-IMX-LMC, Lausanne, Switzerland.

[*] Corresponding author: reza.andalibi@psi.ch

## Abstract

Optimal design of nanoparticle synthesis protocols is often achieved *via* one-at-a-time experimental designs. Aside from covering a limited space for the possible input conditions, these methods neglect possible interaction between different combinations of input factors. This is where mechanistic models embracing various possibilities find importance. By performing global uncertainty/sensitivity analysis (UA/SA), one can map out the various outcomes of the process *vs.* different combinations of operating conditions. Moreover, UA/SA allows for the assessment of the model behavior, an inevitable step in the theoretical understanding of a process. Recently, we developed a coupled thermodynamic-kinetic framework in the form of population balance modelling in order to describe the precipitation of calcium-silicate-hydrate. Besides its relevance in the construction industry, this inorganic nanomaterial offers potential applications in biomedicine, environmental remediation, and catalysis most notably due to ample specific surface area that can be achieved by carefully tuning the synthesis conditions. Here, we apply a global UA/SA to an improved version of our computational model in order to understand the effect of variations in the model parameters and experimental conditions (induced by either uncertainty or tunability) on the properties of the product. With specific surface area of particles as an example, we show that UA/SA identifies the factors whose control would allow a fine-tuning of the desired properties. This way, we can rationalize the proper synthesis protocol before any further attempt to optimize the experimental procedure. This approach is general and can be transferred to other nanoparticle synthesis schemes as well.

## 1. Introduction

Calcium-silicate-hydrate (CaO-SiO$_2$-H$_2$O or C-S-H for short) is the most important phase formed during the hydration of cementitious materials [1]. Aside from its key role in the construction industry [1], C-S-H has recently found diverse applications in environmental clean-up [2–4], biomedicine [5–7], and even catalysis [8,9]. In the biomedical field, for instance, it offers good bioactivity, biocompatibility, and biodegradability [6,7]. Besides these characteristics, the inherent nanostructured construct of C-S-H, provides high surface areas, and its relatively low-cost preparation warrants further research for applications where interfaces play a major role [4,7].

Recently, we developed a formalism to model the nucleation and growth of C-S-H using a population balance equation (PBE) framework [10]. The theoretical framework was fitted to the experimental data collected on the precipitation of a synthetic C-S-H with Ca:Si = 2, prepared under controlled conditions resembling the process of cement hydration (in terms of temporal supersaturation ratio) [10,11]. We estimated the optimal values for the unknown model parameters and explained procedures for the extraction of various output information from the simulation. Additionally, we assessed the merit of our computations by comparing the optimal physical parameters and various outputs against the literature data, wherever available [10].

Here, we build on our previous work and implement two pivotal refinements to improve the simulation speed, robustness, and generality. Specifically, we replace our *ad hoc* equilibrium solver in the previous work with PHREEQC, a popular freely-available tool widely used for thermodynamic speciation calculations [12]. This allows for a more straightforward adaptation to new precipitation scenarios and opens up the possibility of utilizing the large thermodynamic databases already included within the software [12]. Additionally, we employ the direct quadrature method of moments (DQMOM) for the solution of the PBE, which has several advantages over our previously used QMOM approach [13–15]. We give a detailed derivation of DQMOM and relevant subtleties critical to the robust and reliable performance of the method.

Having this improved simulation framework, we assess the behavior of the C-S-H precipitation model by applying global uncertainty/sensitivity analysis (UA/SA) with different model parameters as the source of uncertainty. The propagation of uncertainty into different model outputs such as crystallite dimensions, particle edge length, specific surface areas, and precipitation yield is examined thoroughly using three different methods, namely, PAWN (derived from the developers names, Pianosi and Wagener) [16,17], Elementary Effect Test [18,19], and variance-based sensitivity analysis (VBSA) [19,20]. The application of these complementary methods enables unambiguous appraisal of the model performance, which in turn facilitates complexity reduction, *i.e.*, by fixing uninfluential parameters to reasonable values. This also allows for a more robust calibration during the regression to experimental data [19,21]. Additional implications of such global UA/SA concerns the robustness of model predictions in response to different sources of uncertainty/variability in the model parameters [19,21]. Besides these outcomes, our work provides the first example of UA/SA on a kinetic model of precipitation, and thus can serve as a benchmark for future studies in this direction.

Once we obtained a comprehensive understanding about the model structure, we implement another UA/SA on a model of reduced complexity and incorporate uncertainty from different experimental conditions. The goal is to aid the design of nanoparticulate products by gaining insight from computer experiments. Often, optimal operating conditions for a synthesis protocol are found using one-at-a-time (OAT) experimental designs [7,19]. Besides covering a limited space of the possible input conditions, this practice also overlooks the probable interaction effects between various combinations of the inputs. The latter could produce drastically different behaviour compared to when only one input parameter is changed at a time [19]. A global UA/SA circumvents these limitations and offers an inexpensive alternative to examine a wide range of operating conditions varied in an all-at-a-time fashion [19,21]. With this approach, we propose practical recommendations in order to improve the properties of the final product. As an example, we demonstrate the key influence of reagent addition rate, in a well-mixed semi-batch reactor, on the accessible specific surface area of the final product. This can be further compounded with adjustments in the solution chemistry to obtain a product with distinctly higher specific surface area.

## 2.  Computational Details

The overall computational workflow for PBE modelling of precipitation processes is explained in detail in our recent articles [10,22,23] and other literature [14,24–26]. Therefore, in this section we will focus on the developments brought forward by the current work. First, we will review the essential characteristics of the precipitation system to be studied, and enumerate the limitations of our previous work. Then, we will briefly explain the application of DQMOM to solve the PBE model for a well-mixed system with crystallite size as the internal coordinate (detailed derivations can be found in the Supporting Information (SI) Section 1). Next, we will describe the coupling to PHREEQC. After that, we summarize the overall simulation workflow including the improvements introduced in this work. Finally, we will present the underlying idea behind different sensitivity measures employed to assess the input-output relationships in the overall coupled thermodynamic-kinetic framework. Additional implementation details are presented in the SI Section 1 and 2.

2.1.   Description of Precipitation System and Limitations of Our Previous Work

The precipitation system studied here is the formation of synthetic C-S-H with Ca:Si ratio 2 [10,11]. The precipitate is composed of nanofoils that are a few nm thick and in the order of 100 nm wide (Figure 1(a) and (b)). These two-dimensional nanoparticles are made up of highly defective crystallites, of a thickness typically below 10 nm, which are arranged with liquid crystalline-type orientational order (Figure 1 (b) and (c)). Recently, we proposed a pathway (Figure 1 (d)) for the formation of this nanoparticulate material and tested that by regressing the experimental kinetic data using a computational model based on population balance equation (PBE) modelling. The framework included primary nucleation, true catalytic secondary nucleation, and molecular growth, and accounted for the time evolution of the precipitation driving force by applying thermodynamic equilibrium to the reactions among the aqueous species (local equilibrium assumption [10,14]). From a mathematical perspective, the framework consisted of a set of ordinary differential equations (ODEs) written for the dynamic evolution in the moments of crystallite size distribution (the PBE part) and elemental amounts in the system (the mass balances). Our computational model showed very good plausibility in terms of the goodness of fit, the consistency of the regressed model parameters with respect to the knowledge from the literature, and reasonable mechanistic and kinetic predictions. This includes, for instance, the predicted size of crystallites and particles which were compatible with previous experimental and theoretical observations, or the invariably undersaturated state of solution with respect to portlandite similar to experimental observations [10].

In the computational framework mentioned above, the PBE set was solved using the quadrature method of moments (QMOM) [10,27]. Although QMOM is a reliable and popular method for this task [15,24,28], a widely used variation called direct quadrature method of moments (DQMOM) offers several advantages. For instance, QMOM requires a moment inversion algorithm at every time step to find the discrete approximation to the size distribution. This is often an ill-conditioned problem and reduces the computational efficiency of the method drastically [14,28,29]. On the contrary, DQMOM directly follows the discrete abscissas and weights approximating the size distribution, and employs commonly used numerical methods such as matrix inversion instead of moment inversion algorithms. This makes the method more convergent and extremely fast [13,14,28], to an extent that Haderlein *et al.* [14] have employed it for the development of "flow-sheeting" software tools. Other benefits of DQMOM over QMOM are its more efficient coupling to fluid dynamics and straightforward extension to multivariate distributions (namely, with more than one internal coordinate) [15,29]. The latter is particularly important in the case of C-S-H as oftentimes the precipitation leads to variable composition solid solutions [30] necessitating the application of at least two internal coordinates, namely, size and composition [10].

Another limitation of our previous work is related to the manner of applying the local equilibrium assumption. There, we calculated the supersaturation ratio at fixed time steps selected depending on how fast the precipitation consumes the precursors [10]. Therefore, supersaturation ratio was an externally calculated quantity provided to the ODE function (the function calculating the derivatives as a function of time and dependent variables [31]). This is very similar to the approach adopted by Myerson *et al.* who only recalculated the supersaturation ratio when there was significant (more than 0.1%) change in the amount of the precipitate [26]. This approach was adopted to minimize the computational burden of speciation calculations. Additionally, similar to the work by Haderlein *et al.* [14] and Galbraith and Schneider [32], a bespoke speciation solver was developed to expedite the overall simulation [10]. Even though our equilibrium solver and that by Haderlein *et al.* are developed in a general fashion, this practice limits the applicability of the developed tools, as it requires setting up a thermodynamic database for every single new scenario. Instead, there are a number of powerful freely-available thermodynamic solvers with huge databases already implemented, such as PHREEQC and GEMS [12,33]. Therefore, in this work we will present a general-purpose protocol for the coupling of PHREEQC to the PBE simulations of precipitation processes with the former imbedded within the ODE function.

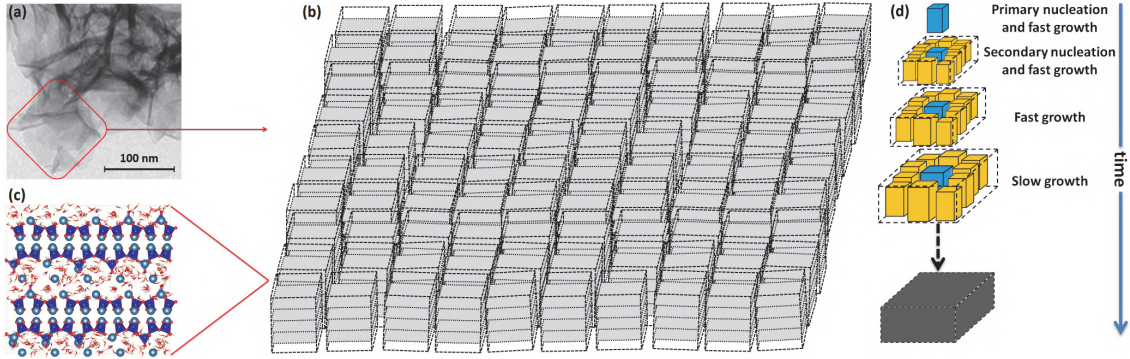

Figure 1. Summary of synthetic C-S-H precipitation system studied here. (a) Transmission electron micrograph of C-S-H particles with foil like morphology; (b) schematic representation of C-S-H nanofoils composed of defective crystallites nematically ordered in two dimensions; (c) internal structure of C-S-H crystallites from atomistic simulations [11,34]; (d) the proposed precipitation pathway for synthetic C-S-H of Ca:Si=2 [10]. Adapted with permission from Ref. [10] Copyright 2018 The Royal Society of Chemistry.

2.2. Population Balance Equation and Its Solution Using DQMOM

For a homogeneous system (namely, one with uniformity across the physical space), the temporal evolution in the number density function (NDF; $n$) of the crystallite characteristic size ($L$ in m) can be expressed using the so-called population balance equation (PBE)

$$\frac{\partial n}{\partial t} = S(L) \tag{1}$$

where $S$ is a source term embracing all the solid formation/transformation processes such as nucleation, growth, and aggregation, inflows and outflows of crystallites, and possible changes in the volume of reaction liquor. In the kinetic modelling of precipitation processes, the PBE is solved along with differential equations written for mass balances (*viz.*, the conservation of elements inside the reactor). In the direct quadrature method of moments, the NDF is approximated by a discretized distribution as

$$n(L_{nm}, t) = \sum_{\alpha=1}^{N} w_\alpha(t) \delta \left( L_{nm} - L_{\alpha,nm}(t) \right) \tag{2}$$

where $\alpha$ denotes the discretization nodes with weights $w_\alpha$ at sizes (or abscissas in nm) $L_{\alpha,nm}$, $N$ is the overall number of the nodes, and $\delta$ is the Dirac delta function [15,29]. Upon substituting Eq. (2) in Eq. (1), and applying the moment transformation ($k$ denotes the moment order)

$$m_k(t) \equiv \int_0^\infty L^k\, n(t,L)\, dL \cong \sum_{\alpha=1}^{N} w_\alpha(t) L_\alpha^k \tag{3}$$

the solution to the population balance equation (Eq. (1)) would amount to solving the following system of linear equations (see SI Section 1 for the complete derivation of all the equations)

$$A_{nm}\alpha_{nm} = [A_{1,nm}, A_{2,nm}]\begin{bmatrix}a\\b_{nm}\end{bmatrix} = d_{nm} \tag{4}$$

with

$$A_{1,nm} = \begin{bmatrix} 1 & \cdots & 1 \\ 0 & \cdots & 0 \\ -L_{1,nm}^2 & \cdots & -L_{N,nm}^2 \\ \vdots & \vdots & \vdots \\ (1-k)L_{1,nm}^k & \cdots & (1-k)L_{N,nm}^k \\ \vdots & \vdots & \vdots \\ 2(1-N)L_{1,nm}^{2N-1} & \cdots & 2(1-N)L_{N,nm}^{2N-1} \end{bmatrix}_{2N \times N} \tag{5}$$

$$A_{2,nm} = \begin{bmatrix} 0 & \cdots & 0 \\ 1 & \cdots & 1 \\ 2L_{1,nm} & \cdots & 2L_{N,nm} \\ \vdots & \vdots & \vdots \\ kL_{1,nm}^{k-1} & \cdots & kL_{N,nm}^{k-1} \\ \vdots & \vdots & \vdots \\ (2N-1)L_{1,nm}^{2N-2} & \cdots & (2N-1)L_{N,nm}^{2N-2} \end{bmatrix}_{2N \times N} \tag{6}$$

$$\alpha_{nm} = [a_1, a_2, \ldots, a_N, b_{1,nm}, b_{2,nm}, \ldots, b_{N,nm}]^T = \begin{bmatrix}a\\b_{nm}\end{bmatrix}_{2N \times 1} \tag{7}$$

$$d_{nm} = [\bar{S}_0, \bar{S}_1 \times 10^9, \ldots \bar{S}_k \times 10^{9k}, \ldots, \bar{S}_{2N-1} \times 10^{9(2N-1)}]^T \tag{8}$$

$$a_\alpha = \frac{\partial w_\alpha}{\partial t} \tag{9}$$

$$b_{\alpha,nm} = \frac{\partial(w_\alpha L_{\alpha,nm})}{\partial t} \tag{10}$$

$$\bar{S}_k \equiv \int_0^\infty L^k S(L) dL \tag{11}$$

Therefore, solving Eq. (4) at each time step of integrating the set of ordinary differential equations (ODE set composed of PBE + mass balances) yields the time derivative terms $a$ and $b_{nm}$ for the weights and weighted abscissas ($w_\alpha L_{\alpha,nm} \equiv \varsigma_{\alpha,nm}$). In the current precipitation system, the source term $d_{nm}$ is composed of contributions from nucleation (primary and secondary; $d_{nm}^N$), molecular growth ($d_{nm}^G$), and changes in the system volume ($d_{nm}^{Volume}$)

$$d_{nm}^N = diag(1, 10^9, \ldots, 10^{9k}, \ldots, 10^{9(2N-1)}) \times \begin{bmatrix} 1 & 1 \\ L_{hom}^* & L_{sec}^* \\ \vdots & \vdots \\ L_{hom}^{*\,k} & L_{sec}^{*\,k} \\ \vdots & \vdots \\ L_{hom}^{*\,2N-1} & L_{sec}^{*\,2N-1} \end{bmatrix} \times \begin{bmatrix} J_{hom} \\ J_{sec} \end{bmatrix} \quad (12)$$

$$d_{nm}^G = diag(1, 10^9, \ldots, 10^{9k}, \ldots, 10^{9(2N-1)}) \times A_2 \times diag(w_1, w_2, \ldots, w_N) \times \begin{bmatrix} G(L_1) \\ G(L_2) \\ \vdots \\ G(L_N) \end{bmatrix} \quad (13)$$

$$d_{nm}^{Volume} = -\frac{d(\ln V)}{dt} diag(1, 10^9, \ldots, 10^{9k}, \ldots, 10^{9(2N-1)}) \times \begin{bmatrix} m_0 \\ m_1 \\ \vdots \\ m_{2N-1} \end{bmatrix} \quad (14)$$

In these equations, $diag()$ denotes a diagonal matrix, $J_{hom}$ and $J_{sec}$ are the rates of primary homogeneous and true secondary nucleation processes (crystallites.m$^{-3}$.s$^{-1}$), $L_{hom}^*$ and $L_{sec}^*$ are the respective critical nuclei sizes (in m), and $V$ is the volume of the reaction suspension.

In the current study, three measures have been exercised to make the DQMOM method more robust. Firstly, the matrix $A_{nm}$ is written with abscissas in nm to reduce the condition number and facilitate the solution of the linear system (that is, Eq. (4) which gives the temporal derivatives of the quadrature weights and weighted abscissas). Secondly, additional reduction in the condition number is attained by left preconditioning (SI Eqs. (26 − 28)) [13,35]. Thirdly, the temporal derivatives of the log$_{10}$ of $w_\alpha$ and $\varsigma_{\alpha,nm}$ are integrated rather than the untransformed variables to bring them into an order of unity and improve the convergence and robustness when using the MATLAB's ODE solver [31].

2.3. Thermodynamic Speciation *via* Coupling to PHREEQC

PHREEQC is a freely-available geochemical reaction solver capable of simulating a variety of processes including solid-liquid-gas equilibria, surface complexation, ion exchange, and much more [12]. Aside from its carefully developed internal database, there are many comprehensive third-party databases including Cemdata18, specifically developed for cementitious systems [30]. With such a broad range of capabilities and extensive thermodynamic infrastructure, coupling PBE simulations with PHREEQC opens up new avenues in the practical and facile application of this powerful method to the understanding and design of particulate processes. Such coupling is facilitated by an already developed module called IPhreeqc, which enables interfacing with different scripting languages such as MATLAB and Python *via* Microsoft COM (component object model) [36,37]. Very recently, a number of articles have been published reporting the coupling of PHREEQC with PBE simulations [38–40]. Nonetheless, to the best of our knowledge none of these publications provided the corresponding computer code and procedures for the implementation.

Here, we developed a function (the file "eqbrmSolver.m" in the SI) that provides a general interface for PHREEQC speciation calculations in MATLAB. Briefly, information about the solution chemistry (*e.g.*, different compounds and their concentrations) and experimental conditions (such as temperature or gases at constant partial pressure in equilibrium with solution) are provided to the interface, which in turn passes the data into PHREEQC solver *via* the COM object. These inputs are provided using keywords in a fashion similar to PHREEQC syntax (Figure S 1). This allows the simulation of precipitation in practically unlimited number of systems and scenarios without the necessity to rewrite the speciation code and/or its database. The information passed as the outputs of speciation calculation are the mass of water solvent, solution density, elemental concentrations, species concentrations and activities, pH, ionic strength, and saturation indices ($SI = \log_{10}(\frac{IAP}{K_{sp}})$ with $IAP$ being the ionic activity product [12]) with respect to different solid phases.

In the current study, we employed the Cemdata18 database [30] for all the aqueous reactions and the density and solubility product of the precipitate (C-S-H with Ca:Si=2) were taken from our previous paper [10].

## 2.4. Overall Simulation Workflow

Figure 2 summarizes the steps involved during the simulation of a precipitation process (implemented in MATLAB R2019a). An example simulation is presented in the "demo.m" script provided in the SI. All the core PBE simulations are handled by "pbe.m" function while the output provided by this function suffices for the calculation of any other output of interest. For instance, knowing the moments allows one to calculate different crystallite and particle characteristics such as size and specific surface area. Similarly, knowing the amount of water solvent (input to "pbe.m") and temporal mole amounts of all the elements, one can back calculate the full speciation using the function "eqbrmSolver.m".

For a typical scenario that simulates 24 hours of precipitation with known model parameters (from our previous work [10]) on an ordinary HP laptop, with dual-core Intel® Core™ i5-4310M CPU @ 2.70 GHz 2.70 GHz processor and 8.00 GB of RAM, the run time was 80 seconds. This is roughly half the time it took using our previous PBE code (which was using QMOM and bespoke speciation solver). It is worth noting that in the current version, in contrast to our previous code, the speciation calculations are embedded within the ODE function. In other words, the calculation is performed at every time step selected by the ODE solver, which makes the total number of such calculations much greater in number than in our previous code. Therefore, the real speed up due to the application of DQMOM (in the modified form introduced here) in place of QMOM is much larger, consistent with previous reports [14,28].

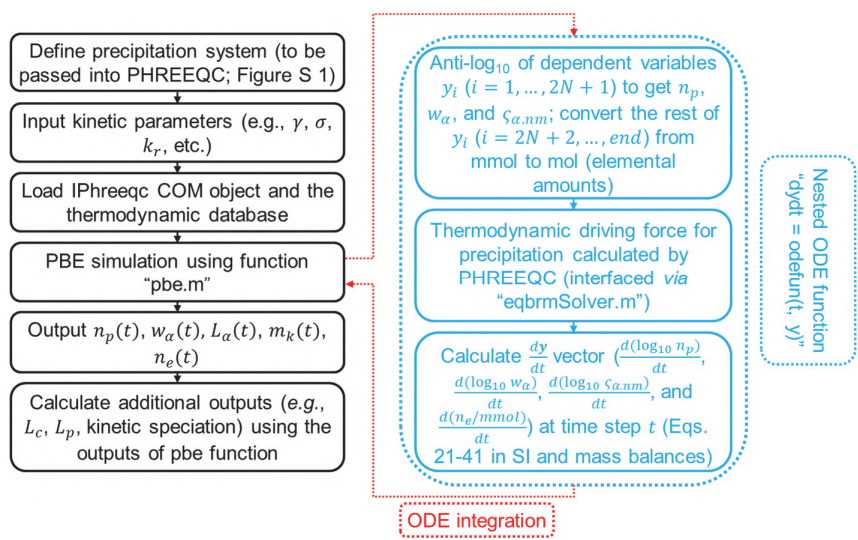

Figure 2. The overall algorithm for the PBE simulation of a precipitation process. Black boxes comprise the main workflow backbone (*e.g.*, "demo.m" script in the SI), blue represents the content of the ODE function, and red refers to the integration of the ODE set using an appropriate solver (MATLAB's ode15s in this case [31]). All the symbols are defined in the main text.

## 2.5. Problem Setting for Uncertainty Assessment

To quantitatively apportion the variability in the output of the PBE simulation to various sources of uncertainty in the input (factors), we applied model-independent sensitivity analysis using three popular methods: PAWN [16,17], Elementary Effect Test or the method of Morris (EET) [18,19], and variance-based sensitivity analysis (VBSA) using quasi-Monte Carlo samples generated by the method of Sobol' [19,20]. The analysis was mostly implemented using the SAFE package, an open-source MATLAB toolbox that includes various functions for the generation of input samples, estimation and evaluation of sensitivity indices, and extensive visualization tools [41]. The target of uncertainty/sensitivity analysis, that is, the PBE model for the precipitation of synthetic C-S-H, has five unknown parameters: interfacial tension ($\gamma$), cohesion energy ($\sigma$), growth rate coefficient ($k_r$), kinetic order of growth ($g$), and crystallite aspect ratio ($a_r$) [10]. Below we will discuss the feasible uncertainty domain for each of these parameters.

The nominal value of interfacial tension from our previous regression to experimental data was estimated to be in the range of 0.05-0.06 J.m$^{-2}$ [10]. Preliminary numerical experiments proved that 0.065 is a suitable upper bound as larger values gave unrealistically hindered nucleation. Under the conditions of interest the surface charge density of C-S-H particles lies in the range of 0.4 C.m$^{-2}$ [42]. From Figure 8.2 in Mersmann's handbook this could bring about 0.025 J.m$^{-2}$ variation in the interfacial tension between C-S-H and liquid solutions. Therefore, a lower bound of 0.04 J.m$^{-2}$ was assumed for $\gamma$. Again, numerical experiments showed that smaller $\gamma$ would result in unrealistically early nucleation (at very low supersaturation ratios) or even spinodal decomposition [43].

Theoretical considerations dictate that the relative cohesion energy between an already precipitated solid substrate and secondary nuclei is in the range 0-2 [10,44–46]. Preliminary tests, however, indicate that in our system of interest values larger than unity would give extremely small effective interfacial tensions for secondary nucleation, particularly when the interfacial tension is close to the lower bound set earlier. Additionally, a value of $\sigma/\gamma$ = 2, which corresponds to coherent interfaces or epitaxial growth, hardly happens in precipitation from liquid solutions because of ions and solvent molecules adsorbed onto the surface of the substrate [46]. This is compounded with the extremely defective nature of C-S-H crystallites hampering the formation of interfaces with matching lattices [10,11]. All things considered, a pragmatic upper bound of 1 was selected for $\sigma/\gamma$.

In our previous work, we fitted values in the order of $10^{-9}$ to the parameter $k_r$ and 2 for parameter $g$ [10]. For the sake of UA/SA, values within one order of magnitude around $10^{-9}$ were considered. As suggested by Marino *et al.* [47], to sample the variability space more uniformly, considering the variation of $k_r$ over two orders of magnitude, $\log_{10} k_r$ was preferred for the sampling. As for kinetic order of growth we sampled in the range 1-3 which is the typical variation range covering rough growth, dislocation-controlled mechanisms and surface nucleation regimes [48,49].

Finally, the ratio of crystallite edge length to its thickness regressed in our previous work was 0.5 [10]. Therefore, an input range 1/3 to 1 was considered for the possible variability. For all the input factors uniform probability distribution functions were considered for the generation of the sample [50].

## 2.6. Sensitivity Measures

In this section, we will briefly explain the three sensitivity methods employed in the current study to facilitate the comprehension of the results. The interested reader is referred to the relevant literature for a more in depth discussion [19,21].

As Saltelli *et al.* [19] argued, UA/SA consists in the examination of uncertainty in parameters (input factors) propagating through a mathematical model all the way to the model outputs [19]. One way to do this task is through Monte Carlo analysis, wherein a set of row vectors are generated by sampling the input variability space of different model parameters. The accumulation of these sets gives an input matrix with each row corresponding to a set of model parameters whose introduction to the model allows a single simulation run. Therefore, any UA/SA requires an input sample matrix.

In this study, low-discrepancy Sobol' sequences were constructed by first generating an input sample $X$ of size $2N \times M$, where $N$ is the *base* sample size, and $M$ denotes the number of uncertain parameters (*e.g.*, $M$ = 5 for five model parameters subject to SA). Sample $X$ was generated using Latin hypercube sampling (LHS) strategy [19,21,41]. This sample was then resampled to build three matrices $X_A$, $X_B$, and $X_C$, where $X_A$ and $X_B$ are simply the first and last $N$ rows of $X$, respectively, while $X_C$ is a block matrix of $M$ recombinations of $X_A$ and $X_B$

$$X_C = \begin{bmatrix} X_{C,1} \\ X_{C,2} \\ \vdots \\ X_{C,M} \end{bmatrix} \quad (15)$$

where $X_{C,i}$ is an $N \times M$ matrix whose columns are all taken from $X_B$ except for the i[th] column which is taken from $X_A$ [19]. Once we have $X_A$, $X_B$, and $X_C$, the PBE model should be run with all the rows in sample matrices as input model parameters giving a total of $N \times (M + 2)$ sample outputs. Here, the size of crystallites and particles, their specific surface areas (SSA), C-S-H precipitation yield (conversion with respect to equilibrium composition), solution pH, and saturation index with respect to portlandite, all quantities after 12 hours of precipitation, are qualitatively examined as model outputs (uncertainty analysis). Subsequently, quantitative assessment in the form of global sensitivity analysis was performed on three selected outputs: crystallite thickness ($\bar{L}_c$), particle edge length ($\bar{L}_p$), and specific surface area of particles ($SSA_p$). These are of special practical relevance and they can be compared to experimentally measured values [2,5,7,10,51]. We started with a base sample size of N = 2000 (corresponding to 14000 input sample points) and extended the $X$ matrix using the SAFE toolbox function "AAT_sampling_extend" to assess the convergence behavior of different sensitivity measures. Throughout this work, bootstrapping over 1000 resamples has been used to estimate the 95% confidence bounds on all the SA indices [41,50].

The first SA method applied here is PAWN, a density-based (or moment-independent) method recently developed by Pianosi and Wagener [16,17]. The central idea behind this method is to compute sensitivity through variations in the cumulative density function (CDF) of the output, induced by fixing one input factor. In practice, this is achieved *via* estimating the divergence between unconditional output CDF, namely that generated by varying all the input factors, and the conditional CDF generated by fixing an individual factor to a prescribed value. Several values within the input variability space can be assigned to the prescribed value, a practice referred to as multiple conditioning, to generate a number of divergence values that can be aggregated in some kind of statistic [16,17,21]. In PAWN, the divergence is expressed in terms of Kolmogorov-Smirnov (KS) statistics which is the maximum vertical distance between the conditional and unconditional CDFs [16,17]. PAWN, and moment-independent methods in general, are particularly useful in case of highly skewed or multimodal output distributions. In such cases, variance is not an adequate proxy of uncertainty and variance-based methods (see below) can no longer be applied [17]. Another advantage of these methods is that they can be estimated from generic samples, that is, without requiring tailored sampling strategies [17,21]. Therefore, in this work we use the samples generated as described earlier to estimate the average and maximum of KS statistics calculated over 10 conditioning intervals [50]. To distinguish influential and uninfluential input factors, following Khorashadi Zadeh *et al.* [52] we artificially introduced a dummy input factor that does not appear in the model and, thus has no impact on the output. Therefore, the sensitivity index (maximum of KS statistic) corresponding to this factor defines the threshold for parameter screening [50,52].

The second SA method used here is the Elementary Effect Test [18,19]. In this case, the idea is to correlate model sensitivity with the effect of perturbing the input factors—one at a time—on the model output. An example of this approach is to estimate (*e.g.*, by finite differences) the partial derivatives with respect to different model parameters at their nominal values. In this form, the method is computationally very cheap but only provides local sensitivity information [21]. A global extension of this technique is to compute perturbations from multiple points within the input variability space, followed by aggregating them in some type of statistic. The most popular method in this group uses the average of $r$ finite differences (also known as Elementary Effects or EEs) as the sensitivity measure ($\mu_{EET}$) [19,21]. Here, a refined measure taking the absolute values of EEs is used to avoid cancelation due to sign differences [53]. Beside the average of EEs, standard deviations ($\sigma_{EET}$) provide information about the degree of interaction between the parameters and/or their level of nonlinearity [19,21]. To apply EET using the Sobol' sample we discussed earlier, the input and output were converted to the format required by EET functions. This was done using the function "fromVBSAtoEET" in the SAFE toolbox, which rearranges the matrices $X_B, Y_B, X_C$, and $Y_C$ so that they can be used to calculate EET indices from a radial design [19,41]. With this approach, the number of sampling points ($r$) would be equal to the Sobol' base sample size ($N$).

The last method employed in the current study is the variance-based SA (VBSA) [19,20]. This method assumes that the output variance is an indicator of its uncertainty and the contribution of each input factor to this variance is a measure of sensitivity. This technique handles nonlinear and non-monotonic functions/computational models, as well as those exhibiting interactions between their factors. Besides, it is able to capture the influence of each factor's full-range of variation [20,54,55]. Perhaps, the biggest drawback of this method is the large number of simulation runs it requires for convergence [21,55]. Aside from computational aspects, another limitation of VBSA is that, variance is not a meaningful gauge for highly skewed or multimodal output distributions and hence, for such situations VBSA indices are not appropriate measures of sensitivity anymore [17,21].

In VBSA, typically two types of indices are defined, first-order and total-order. First-order indices (also known as main effects, $S_i$) measure the direct contribution of individual input factors to the variance of output distribution. Equivalently, this can be thought of as the reduction in the output variance achievable by fixing inputs one at a time [56]. In a model where output variability is only a result of main effects (lack of interactions between inputs), $\sum_{i=1}^{M} S_i = 1$ and the model is said to be additive. Nevertheless, in complex computational models, this is rarely the case and main effects do not sufficiently describe the output variability. Considering the high computational expense, particularly for larger $M$, that has to be incurred to estimate all the interaction effects, one may calculate total-order sensitivity indices, $S_{Ti}$, which embrace the main effect as well as all the interactions (of any order) involving the input factor $x_i$ [56,57]. Considering the nature of total effect indices, they are particularly suited for parameter screening as having zero total effect is a necessary and sufficient condition for an input parameter to be uninfluential [21].

## 3. Results and Discussion

### 3.1. Uncertainty Analysis with Model Parameters as Input Factors

In this section, we employ visual tools (scatter plots and histograms) to appraise the propagation of uncertainty from model parameters to different model outputs. Figure 3(a) shows the histograms for average crystallite thickness and edge length ($\bar{L}_c$ and $a_r\bar{L}_c$, respectively) with probability normalization while Figure 3(b) portrays the corresponding results for particle edge length ($\bar{L}_p$). From Figure 3(a) we can see that the crystallite thickness and edge length are typically a few nm, consistent with previous reports for different C-S-H products [7,10,11,51,58,59]. Particle edge length, instead, is typically 1-2 orders of magnitude larger and its distribution assumes a very long tail spanning up to a few μm (Figure 3(b)) [60].

Concerning the specific surface area (*viz.*, surface area per unit mass of precipitate), the hierarchical structure of the solid C-S-H gives rise to two types of surfaces. One is the overall area of the external surfaces of crystallites (*i.e.*, neglecting the internal crystallite structure; $SSA_{Crystallite}$), and the other only considers the external surface of particles neglecting the surfaces embedded within the bulk of the particles ($SSA_{Particle}$). Therefore, by definition $SSA_{Particle} \leq SSA_{Crystallite}$ with the equality happening in the absence of secondary nucleation and aggregation (in other words, when every crystallite is a particle by itself). $SSA_{Crystallite}$ is calculated from the zeroth moment of crystallite size distribution (which is directly available from PBE simulations) and $\bar{L}_c$ while we can estimate $SSA_{Particle}$ from $n_p$ and $\bar{L}_p$ (see the Electronic Supplementary Information in Ref. [10] for the estimation of $\bar{L}_p$ using geometrical considerations). Figure 3(c) shows the scatter plots for $SSA_{Particle}$ *vs.* $SSA_{Crystallite}$ and Figure 3(d) provides the corresponding histograms. In Figure 3(c), we can easily see that the condition $SSA_{Particle} \leq SSA_{Crystallite}$ holds (no data above the identity line) confirming the correct performance of our PBE simulations. Another important constraint is that $SSA_{Crystallite}$ has to always be smaller than or equal to the total surface area of C-S-H building units. Assuming all the calcium in solution is converted to C-S-H and knowing the surface area of a single building unit (1.1297×10$^{-18}$ m$^2$; estimated from C-S-H molar volume [10] and considering a cubic shape, which has the highest surface area among cuboids with similar volume), the theoretical upper bound for $SSA_{Crystallite}$ would be 6014 m$^2$/g. As we can see in Figure 3(c) we are well within this constraint. From Figure 3(d), with the presumed input variability space, the typical values for $SSA_{Crystallite}$ and $SSA_{Particle}$ are a few tens up to around 2000 m$^2$/g.

Another output of PBE simulation is the precipitation yield calculated as

$$Precipitation\ yield\ (\%) = \frac{V_0 \times c_{Ca,0} - V_{end} \times c_{Ca,end}}{V_{end} \times c_{Ca,SLE}} \times 100 \qquad (16)$$

where $V_0$ and $V_{end}$ refer to the initial and final volume of the reaction medium, respectively, $c_{Ca,0}$ and $c_{Ca,end}$ are the corresponding Ca concentrations, and $c_{Ca,SLE}$ is the Ca concentration at the end if precipitation reaches equilibrium. From Figure 3(e), there is a strong relation between $\gamma$ and precipitation yield, wherein the yield can drop significantly at higher $\gamma$. This is anticipated because larger $\gamma$ hinders the onset of primary nucleation [10]. It is worth noting that much weaker correlations, if at all, are observed with other input parameters (Figure S 2).

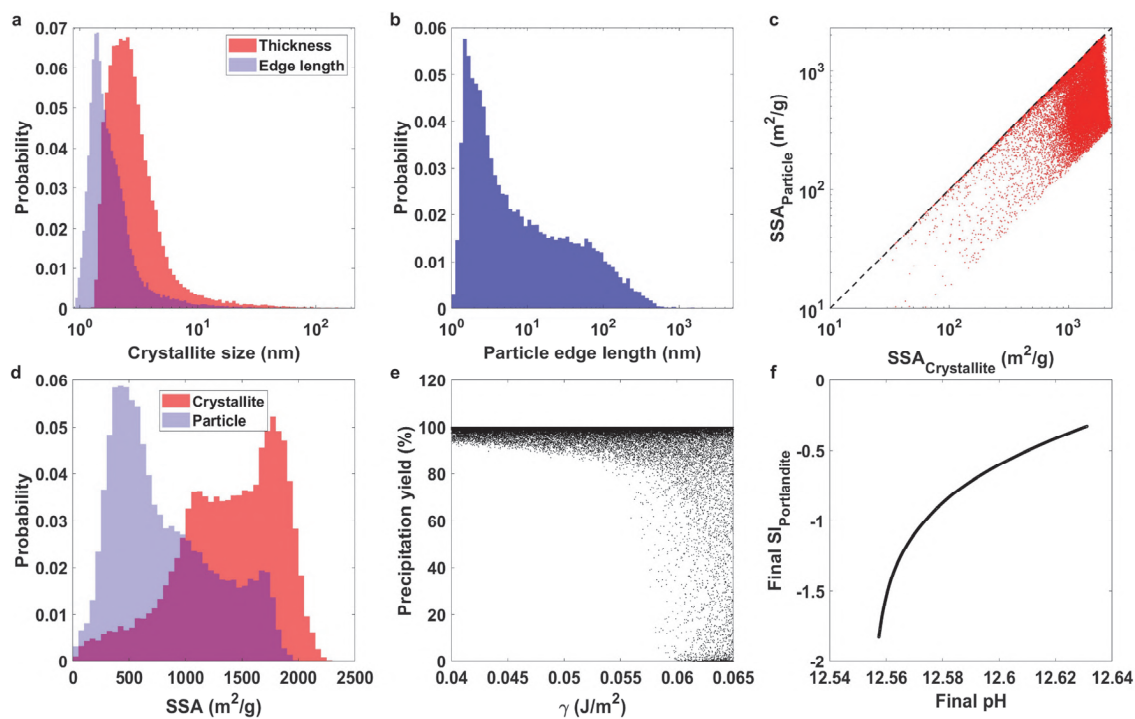

Figure 3. Results of uncertainty analysis with model parameters as input factors. (a) Probability-normalized histograms of crystallite thickness and edge length; (b) probability-normalized histogram of particle edge length; (c) specific surface area (SSA) of particles *vs.* that of crystallites (the black dashed line is the identity line); (d) probability-normalized histograms of crystallite and particle SSA; (e) scatter plot of precipitation yield *vs.* interfacial tension; (f) dependence of final saturation index with respect to portlandite on final solution pH.

Another simulation output is the saturation index ($SI$) with respect to portlandite, a solid phase that competes with C-S-H for precursor ions during the precipitation [10,11,51]. In Figure 3(f) we have plotted this quantity *vs.* the solution pH at the end of the precipitation (after 12 h). From this plot, an unambiguous correlation is visible, where $SI$ increases monotonically with pH. In our previous work, we showed that under the examined operating conditions the system was always undersaturated with respect to portlandite, consistent with experimental observations [10,11]. Nevertheless, according to Figure 3(f) at higher pH values portlandite may precipitate along with C-S-H. Indeed, our simulations with nominal model parameters [10] but at a higher inflow NaOH concentration (*e.g.*, 4 times the value reported in [10] giving a final pH of 13.2; Figure S 14(a)) showed that the system does become supersaturated with respect to portlandite in line with certain experiments (refer to SI Section 3 for further discussion) [51,61].

Now let us examine the mapping of input uncertainty to three selected outputs $\bar{L}_c$, $\bar{L}_p$, and $SSA_p$. From Figure 4, we see that variability in parameter $g$ has almost no effect on any of the outputs (as implied by the uniformity of the scattered points and the lack of pattern [21]). Concerning other input factors, however, the relative degree of uncertainty propagation depends on the output. From Figure 4 (a-c, e), parameters $\gamma$, $\sigma/\gamma$, $\log_{10} k_r$, and $a_r$ are all influential with respect to $\bar{L}_c$ as the output, with $\sigma/\gamma$ having less impact compared to others. Consulting Figure 4(f-h, j), uncertainty in $\sigma/\gamma$ clearly has the highest effect on the variability of $\bar{L}_p$ (strong pattern formed by the scattered data points) while $\gamma$, $\log_{10} k_r$, and $a_r$ have much less of an impact. A similar argument applies to $SSA_p$ (although to a lesser extent) with the output being much more sensitive to $\sigma/\gamma$ (Figure 4(k-m, o)). In the next section, we will present the quantitative assessment of sensitivity with respect to different model parameters.

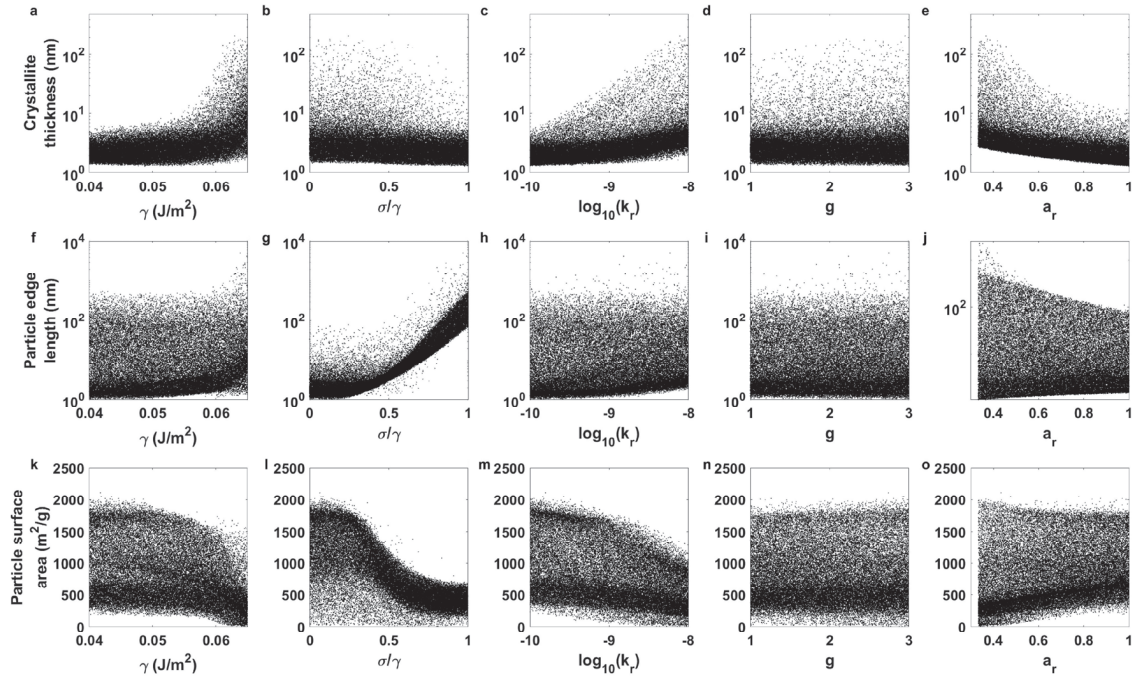

Figure 4. Scatter plots of crystallite thickness (a-e), particle edge length (f-j), and particle surface area (k-o) *vs.* different input model parameters (base sample size 20,000).

### 3.2. Sensitivity Analysis with Model Parameters as Input Factors

Figure 5 summarizes the PAWN sensitivity indices calculated with $\bar{L}_c$, $\bar{L}_p$, and $SSA_p$ as the outputs (obtained from a sample of size 140,000; see SI Section 3 for the convergence analysis of the indices). The top row is the mean KS statistic across the ten conditioning intervals while the bottom row present the maximum of KS. The latter can also be used to identify the influential and uninfluential input factors by comparing the sensitivity indices to that of a dummy variable (which has no effect on the model outcome) [50,52]. From both indices, we can clearly verify the minimal effect of $g$ on the studied model outputs consistent with the conclusions made from the scatter plots as discussed earlier (see previous section and Figure 4). Indeed, taking the maximum KS statistic as the sensitivity measure, it is barely higher than the value estimated for the dummy input (Figure 5(d-f)). With $\bar{L}_c$ as the model output, the rest of the parameters are all influential with $\sigma/\gamma$ being slightly above the dummy variable, and $\gamma$, $\log_{10} k_r$, and $a_r$ exhibiting quite similar higher influences (Figure 5(a,d)). With $\bar{L}_p$ and to a lesser extent $SSA_p$, variability in $\sigma/\gamma$ has the highest impact on the output uncertainty. Except for the uninfluential factor $g$, the rest of model parameters have similar impact over these two outputs (Figure 5(b,c,e,f)).

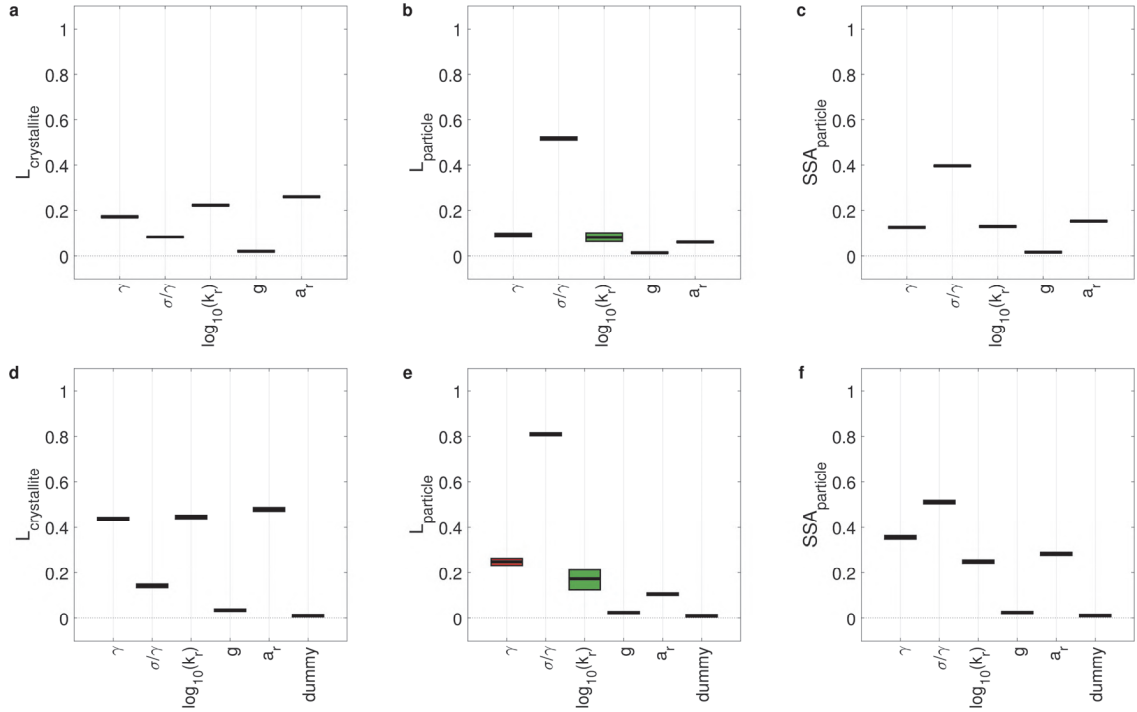

Figure 5. PAWN sensitivity indices in the form of mean (a-c) and maximum (d-f) of the KS statistic, with 95% confidence intervals obtained from bootstrapping, for crystallite thickness (a,d), particle edge length (b,e), and particle surface area (c,f) as the outputs (sample size of 140,000).

Figure 6 summarizes the SA results using the method of Morris (see SI Section 3 for the convergence analysis of the indices). We have plotted the average of absolute values for the EEs ($\mu_{EET}$) against the standard deviations of EEs normalized by their respective averages ($\sigma_{EET}/\mu_{EET}$). We will refer to the latter as the coefficient of variation (C.V.) although strictly speaking C.V. is obtained using the average of signed EEs (and not the absolute values) [18,19,62]. Again, consistent with the results obtained from the scatter plots (Figure 4) and PAWN (Figure 5), we observe very little effect from $g$ on different outputs (Figure 6; $\mu_{EET}$ for $g$ is invariably much smaller than that of the most influential factor). With $\bar{L}_c$ as the model output, $\sigma/\gamma$ is identified as the second least influential parameter while close values are predicted for the other inputs (similar to PAWN; Figure 6(a) and Figure 5 (a,d)). Along the same lines, with $\bar{L}_p$ and to a smaller degree $SSA_p$, $\sigma/\gamma$ has the highest impact on the output with sensitivity indices being 48 and 12 times that of $g$, respectively (Figure 6(b,c)).

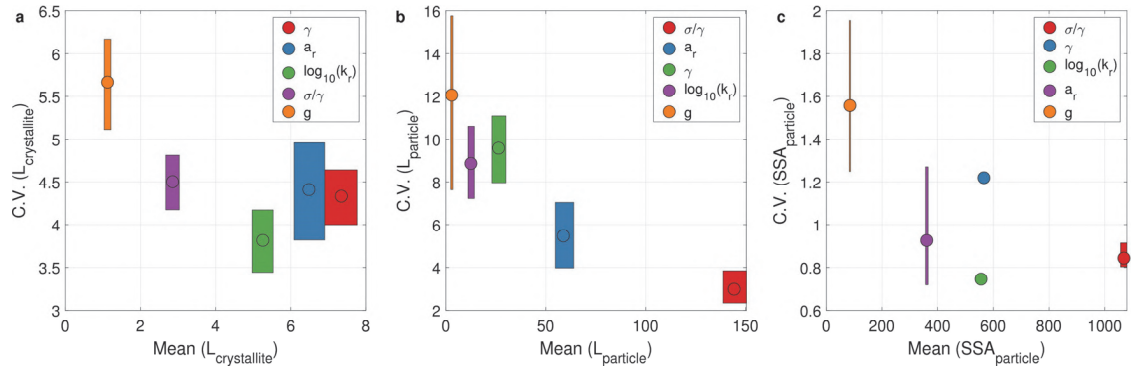

Figure 6. Sensitivity indices obtained using EET with crystallite thickness (a), particle edge length (b), and particle surface area (c) as the outputs (sample size of 120,000). The results are presented as the mean of Elementary Effects plotted against their coefficients of variation (all the 95% confidence intervals are estimated by bootstrapping).

Another observation from our EET analysis is related to the level of nonlinearity in model parameters and/or the degree of interaction between them. Following a method proposed by Garcia Sanchez et al. [63], C.V. values can be classified in four regions [0, 0.1], [0.1, 0.5], [0.5, 1], and > 1. These regions correspond to almost linear, monotonic, almost monotonic, and markedly non-monotonic and/or interacting parameters, respectively. From Figure 6(a,b) (and Figure S 5(a,b)), we see that with $\bar{L}_c$ and $\bar{L}_p$ as outputs all the parameters exhibit a high degree of nonmonotonicity and/or interaction. With $SSA_p$, however, although parameters $\gamma$ and $g$ exhibit highly nonlinear and/or interactive behavior, the output is almost monotonic with respect to parameters $\sigma/\gamma$, $\log_{10} k_r$, and $a_r$ (Figure 6(c) and Figure S 5(c)).

Figure 7 summarizes the SA results based on the variance-based method of Sobol'. Here, considering the sluggish convergence of the sensitivity measures (in particular, the total effects for $\bar{L}_c$ and $\bar{L}_p$, Figure S 6(d,e); see SI Section 3 for the convergence analysis of the indices), we have further explored the possible application of rank (Figure 7(d-f)) and $\log_{10}$ (Figure 7 (g-i)) transformations which have been very popular in the literature [57,64].

From Figure 7(a) we see that with $\bar{L}_c$ as the output the main effects are generally small, with $\sigma/\gamma$ and $g$ assuming zero and $\log_{10} k_r$ being barely above zero. On the other hand, the total effects account for the output variability signifying the nonadditive nature of the input factors [19,20,64]. This is consistent with the C.V. values observed from EET (Figure 6(a) and Figure S 5(a)) all being larger than 1. Nevertheless, even at such a high sample size (140,000) the confidence intervals are wide and there is overlap between the total effect of $\sigma/\gamma$ with $\log_{10} k_r$, and $\log_{10} k_r$ with $g$ (Figure 7(a)). Therefore, we attempted a second SA fixing $g = 2$ (which we already know is practically uninfluential) and going up to a sample size of 288,000 (base sample = 48,000). Doing that, the confidence intervals of $\sigma/\gamma$ fall well below the other three influential parameters ($\gamma$, $\log_{10} k_r$, and $a_r$; Figure S 7(a)). This outcome is consistent with our previous results from PAWN and EET (Figure 5(a,d), Figure 6(a), and Figure S 5(a)).

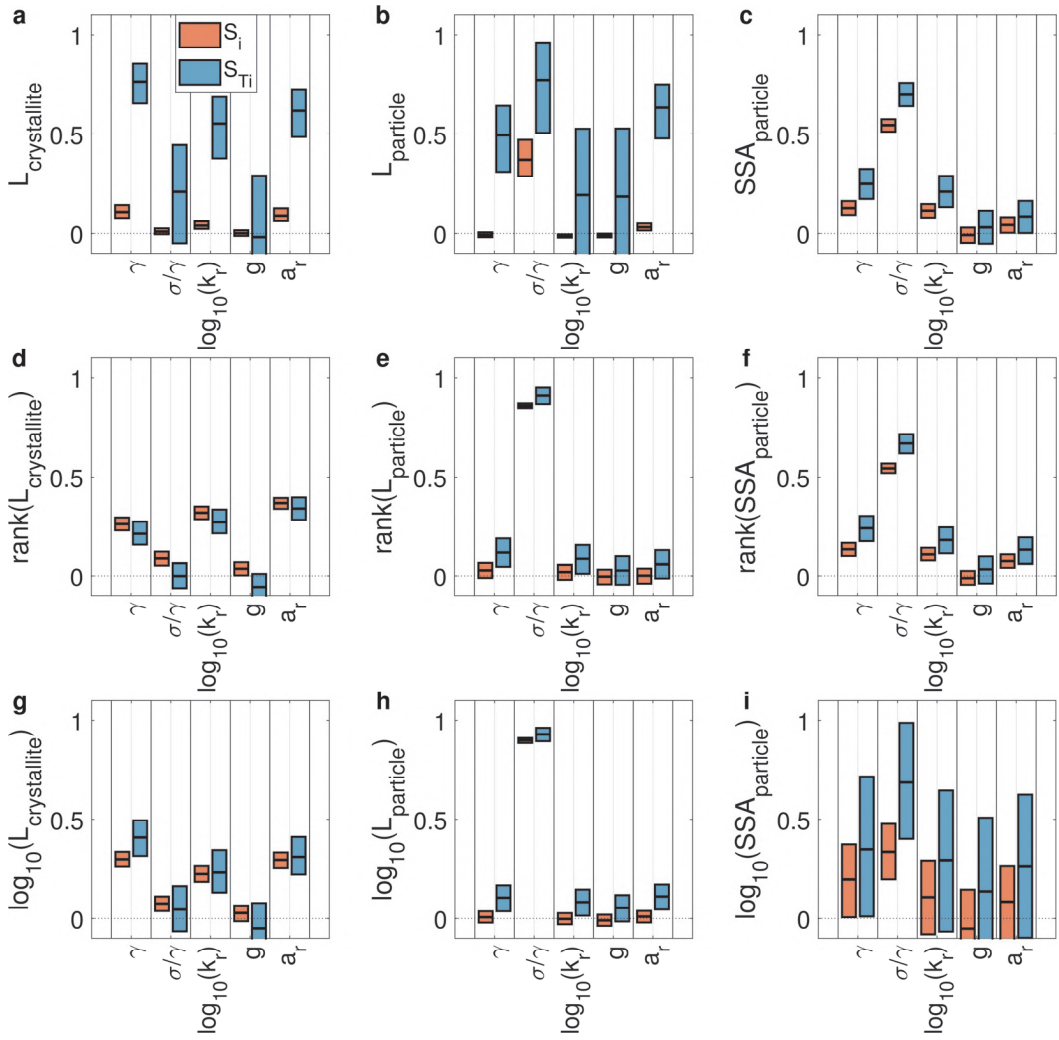

Figure 7. Variance-based sensitivity indices for $\bar{L}_c$, $\bar{L}_p$, and $SSA_p$ (a-c) and their rank (d-f) and $\log_{10}$ transformations (g-i) as the outputs, with a sample size 140,000 (base sample size $N$ = 20,000; $S_i$ and $S_{Ti}$ are the main and total effects, respectively, and the 95% confidence intervals are estimated by bootstrapping).

Consulting Figure 7(b), we note the same problem with $\bar{L}_p$ as with $\bar{L}_c$. This time, however, even at an input sample size of 288,000 the confidence bounds overlap significantly (Figure S 7(b)). This observation can readily be explained by looking at the probability histograms of the output, extending over three orders of magnitude (Figure 3(b)). In other words, the complication arises from the highly skewed distribution of $\bar{L}_p$ with a skewness of 30 (compare with 10.7 for $\bar{L}_c$; Figure 3(a)) [17,21]. One remedy to this problem is the application of rank (Figure 7(e) and Figure S 7(e)) or $\log_{10}$ (Figure 7(h) and Figure S 7(h)) transformation both making the confidence intervals very narrow. Nonetheless, we have to note that sensitivity analysis on transformations of an output cannot always be converted back the non-transformed output [57,64,65]. In fact, rank transformation frequently increases the relative weight of the main effects at the expense of interaction terms. Consequently, the effect of those factors influencing the output mostly by way of interaction with other parameters may be underestimated in a rank-based analysis [64]. In a similar way, in our particular example (Figure 7(d-i) and Figure S 7(d-i)) for all the transformed outputs (both rank and $\log_{10}$) the interactions are dampened rendering the total effects almost equal to the main effects. Especially, with parameters $\gamma$, $\log_{10} k_r$, and $a_r$ affecting $\bar{L}_p$ mainly *via* interactions (see their zero main effects in Figure 7(b) and Figure S 7(b)), upon transformation they apparently become much less influential (that is, they adopt smaller total effects; Figure 7(e,h) and Figure S 7(e,h)). The same complication can be traced in Figure 7(d,g) (and Figure S 7(d,g)) because parameters mainly affect $\bar{L}_c$ by way of interactions.

Another interesting feature can be seen in variance-based indices with $\log_{10} SSA_p$ as the output (Figure 7(i) and Figure S 7(i)). Here, in contrast to the case with untransformed output variable (Figure 7(c) and Figure S 7(c)), the indices have very broad confidence intervals that significantly overlap and make any conclusive deduction impossible [21]. A closer examination of probability distributions of $SSA_p$ and its $\log_{10}$ transformation reveals that while the former is almost unimodal (Figure 3(d)) the latter is highly multimodal (Figure S 8). Quantitatively, the Hartigan's dip test of unimodality [66,67] gives p-values of 0.11 (insignificant multimodality) and 0 (significant multimodality) for the untransformed and $\log_{10}$ transformed variable, respectively. Therefore, aside from the complication in converting SA results from transformed outputs back to the original ones [57,64,65], $\log_{10}$ transformation may render the output distribution multimodal limiting the applicability of VBSA to such scenarios [16,21]. It is worth noting that the latter problem should not happen with rank transformation, as the converted distribution is always uniform and thus unimodal [68].

### 3.3. UA/SA with Selected Model Parameters and Experimental Conditions as Input Factors

Now that we have examined the model behavior in detail, we can turn our attention to the holy grail of the current study, that is, the theory-driven design of nanoparticle synthesis processes. Ideally, a precipitation model should be able to explain the process as a function of experimental conditions alone. In other words, all the parameters in the theoretical framework have to be defined as a function of operating conditions such as temperature, concentrations of reagents, ionic strength, *etc*. This is not an easy task because the development of such models requires extensive and sometimes independent sets of experimental data to identify the mechanistic steps involved and calibrate the corresponding theoretical constructs. For instance, Schroeder *et al.* attempted to calibrate such a framework for the formation and polymorphic transformation of calcium carbonate [24]. Although they accounted for different physicochemical aspects and correlated different parameters with the environmental conditions inside the reactor, limited success was achieved in reproducing the experimental data given the extremely complicated nature of the precipitation process. In the specific case of C-S-H precipitation, additional complications arise due to the nature of the precipitate usually forming a solid solution whose composition depends on the environmental conditions and may evolve as a function of time [10,69,70]. Therefore, with the experimental kinetic data being scarce for synthetic C-S-H [10,11], it is only possible to semi-quantitatively design the product properties as we will present in this section.

In its novel environmental [2–4], biomedical [5–7], and catalysis applications [8,9], the accessible specific surface area of C-S-H product (*i.e.*, $SSA_p$) is one of the most important properties of interest. Therefore, in this section we mainly focus on this characteristic, while information about crystallite and particle sizes are addressed for benchmarking against the literature data.

From the discussion in the previous sections, we found that among the model parameters $g$ is significantly less influential and its impact is barely above the dummy variable. Additionally, our previous studies showed that the aspect ratio of C-S-H crystallites is 0.5 irrespective of mixing flow rate [10]. Interestingly, the same aspect ratio was found for lower Ca:Si solids based on atomistic simulations [71]. Therefore, in this section we fix these parameters to nominal values $g = 2$ and $a_r = 0.5$. The rest of the model parameters were also constrained within reasonable neighborhoods of regressed parameters taken from our previous study ($0.05 \leq \gamma \leq 0.06; 0.5 \leq \sigma/\gamma \leq 1; -9 \leq \log_{10} k_r \leq \log_{10}(5 \times 10^{-9})$) [10]. This procedure allows us to account for potential variations in these parameters when the experimental conditions deviate from those under which the regression data were collected. Having the uncertainty window for the model parameters in place, we investigate the effect of four experimental variables on the outcome of precipitation, in a moderately wide neighborhood of the nominal experimental conditions we previously used for regression [10]. This includes the precipitation temperature (10-50°C), the initial concentration of $Ca(NO3)_2$ inside the reactor (0.01-0.1 mol/kg water; the ratio of $Na_2SiO_3$ to $Ca(NO3)_2$ is kept constant at the original value of 0.5), the concentration of NaOH in the inflow stream (0.1-0.4 mol/kg water), and the rate of inflow stream ($Q$; 0.5-10 mL/min).

Figure 8 summarizes the distributions of crystallite and particle sizes and specific surface areas. With the model parameters varying in a practically more accessible neighborhood of the values regressed to experimental data, we anticipate the distributions presented in this figure match the variations in real systems more closely. As before, the crystallite dimensions are in the range of a few nm, in agreement with experimental observations for different C-S-H products [7,10,11,51,58,59] (Figure 8(a)). It is worth noting that the extreme values of C-S-H crystallite size can be obtained experimentally by carefully adjusting the synthesis conditions. For instance, Mehrali *et al.* obtained C-S-H crystallites as large as 13 and 25 nm in the presence of sodium dodecyl sulfate [72].

From Figure 8(b) we see that the particle edge length can assume values up to a few hundreds of nm. It is worth mentioning that there are fewer reports on the size of synthetic C-S-H particles, which may coincide with the correlation/cutoff length commonly measured in SANS and SAXS and is of the same order as what we observe here [73–75]. TEM images also give values within the range of a few tens to a few hundreds of nm for the width of C-S-H particle [10,11,72].

From Figure 8(c) the condition SSA$_{Particle}$ ≤ SSA$_{Crystallite}$ can be verified. In fact, with the lower bound for $\sigma/\gamma$ being 0.5, there is always significant contribution from secondary nucleation rendering the SSA$_{Particle}$ smaller than SSA$_{Crystallite}$. From Figure 8(d), we note that the corresponding distributions for SSA$_{Particle}$ and SSA$_{Crystallite}$ give values in the order of 400 and 1600 m$^2$/g.

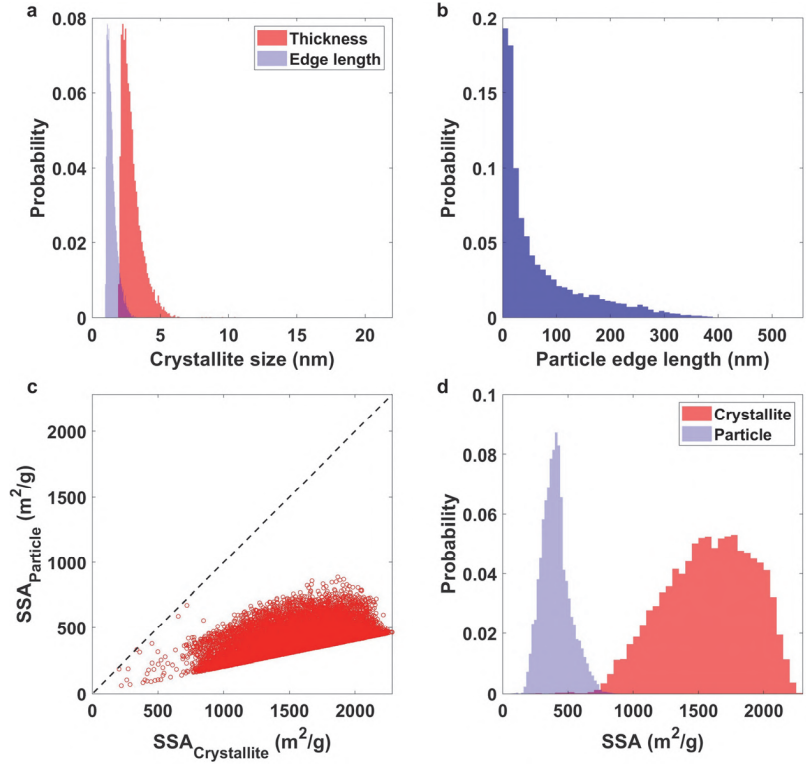

Figure 8. Results of UA with selected model parameters and experimental conditions as the input factors. (a) Probability-normalized histograms of crystallite thickness and edge length; (b) Probability-normalized histogram of particle edge length; (c) Specific surface area (SSA) of particles *vs.* crystallites (the black dashed line is the identity line); (d) Probability-normalized histograms of crystallite and particle SSA.

Figure 9(a-g) presents the scatter plots for $SSA_p$ as a function of different individual input factors. Among the model parameters, $\sigma/\gamma$ appears to be the most influential factor, in line with our results in the previous sections (note the pattern formation in Figure 9(b)). Among the experimental conditions, the addition flow rate of silicate solution seems to dominate the output variability, albeit with a lower impact when compared to $\sigma/\gamma$ (Figure 9(g)). Figure 9(h) shows the colored scatter plot for these two factors with marker colors proportional to the output value. The emergence of color patterns in such a plot is a simple and intuitive tool to assess the degree of interaction between pairs of input factors [21,41]. From Figure 9(h) a weak pattern can be discerned (upper left region) where simultaneous occurrence of high $Q$ and low $\sigma/\gamma$ gives rise to exceptionally higher surface areas (see also Figure S 11(c) presenting the corresponding EET results where C.V. for the three most influential parameters are all below unity indicating weak interactions among the parameters).

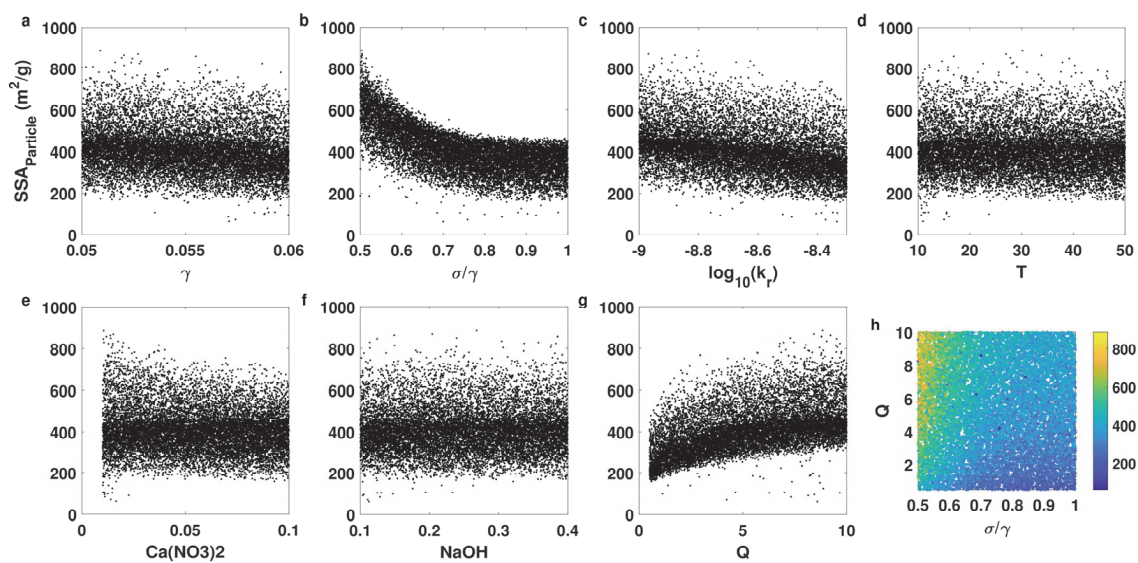

Figure 9. Scatter plots of particle surface area *vs.* different factors in UA with selected model parameters and experimental conditions as the uncertain inputs. (a-g) Output plotted against individual inputs; (h) Scatter plot of $Q$ *vs.* $\sigma/\gamma$ with marker color proportional to the value of output.

For a quantitative assessment of variability propagation to model outputs, PAWN sensitivity indices were estimated for different model parameters and experimental conditions as input factors. Figure 10 summarizes the results with $\bar{L}_c$, $\bar{L}_p$, $SSA_p$ as the outputs (consult Figure S 10 for the convergence of PAWN indices; similar conclusions can also be obtained using EET as depicted in Figure S 11). For $\bar{L}_c$, $\log_{10} k_r$ and $Q$ are the most influential factors (Figure 10(a,d)). All of the experimental variables have low influences on $\bar{L}_c$ barely above the dummy index (Figure 10(d)). The larger impact of flow rate can be understood from the fact that at higher addition rates, the supersaturation build up is larger which in turn induces more contribution from nucleation events to the overall precipitate. Put differently, higher nucleation rates give rise to larger number of crystallites among which the remaining precursor is divided giving rise to smaller crystallites (the same trend was also detected in our previous work; see Table 1 in Ref. [10]).

Looking at Figure 10(b,e), $\bar{L}_p$ is most sensitive to $\sigma/\gamma$ with the rest of input factors having minimal effects only marginally above the dummy index. Physically, this means that the relative rates of primary and secondary nucleation events determine the final particle size.

With $SSA_p$ as the SA target, again $\sigma/\gamma$ is the most influential parameter (Figure 10(c,f)) compatible with our scatter plots (Figure 9(b)). Besides, among the experimental conditions, we can distinguish a comparable dependence on $Q$ (Figure 10(c,f); similar inference as in scatter plot Figure 9(g)). Conversely, the $SSA_p$ of the product is much less sensitive (in a global sense) with respect to the other experimental variables. This is a favorable outcome as it allows for optimizing this key property by tuning the synthesis conditions. Therefore, higher surface areas can generally be obtained by increasing $Q$ irrespective of the value other uncertain input factors assume. From a physical point of view, this can again be explained in the light of supersaturation buildup brought about by higher $Q$ (providing the limiting reactants—$Na_2SiO_3$ and NaOH—faster), which favors primary nucleation over secondary nucleation (and nucleation, in general, over growth) [76]. Consequently, higher particle number concentrations are obtained making the overall surface area larger.

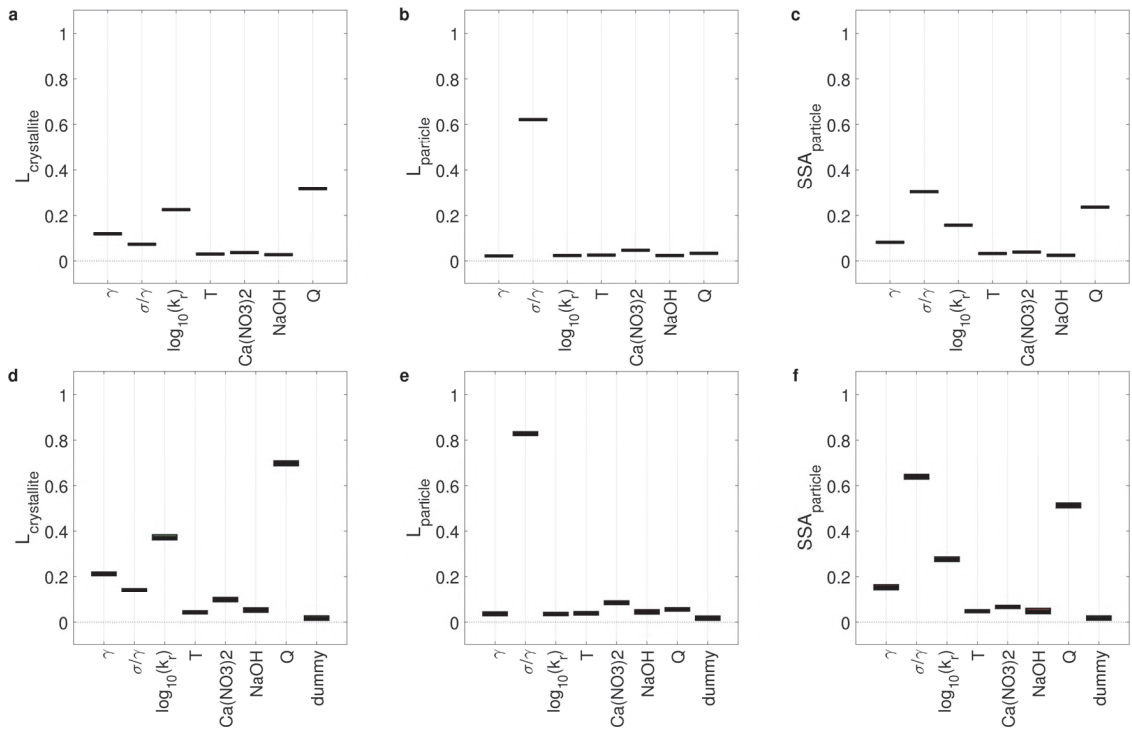

Figure 10. Results of SA with selected model parameters and experimental conditions as the input factors. PAWN sensitivity indices in the form of mean (a-c) and maximum (d-f) of KS statistic, with 95% confidence intervals obtained from bootstrapping, for crystallite thickness (a,d), particle edge length (b,e), and particle surface area (c,f) as the outputs (sample size of 54,000).

Previously, Wu *et al.* synthesized C-S-H (of lower Ca:Si ratios) with specific surface areas ranging between 100 and 500 m$^2$/g, obtained by varying the synthesis conditions [7]. Our results show that there is room for further improvement by increasing the addition flow rate of silicate solution although one has to optimize the design of the synthesis reactor for maximal mixing [77]. This can be reinforced by the synergistic effect of lowering the cohesion energy (that is, lowering $\sigma/\gamma$; Figure 9(b)) which can be induced either by increasing the relative concentration of monovalent ions (*e.g.*, adding a sodium salt to the mixture) or by working at lower pH values [42,78]. Of course, this conclusion only applies to the set of synthesis conditions investigated here and carefully calibrated computational models are needed to cover scenarios that are more diverse. This includes, for instance, different reactant ratios (which typically induces variations in the Ca:Si ratio of the precipitate [9]), alternative addition orders, different pH levels, and the inclusion of other reagents/surfactants (beside those used in the original experiments [10]).

## 4. Conclusions

In summary, we presented a faster, more user-friendly, and more robust version of our previous PBE modelling framework (put forward in Ref. [10]) describing the process of precipitation from liquid solutions.

This was achieved by replacing our speciation function with an interface to PHREEQC, providing access to the large databases already implemented in this popular software. Thanks to this modification, the adaptation to new precipitation scenarios is made much more straightforward and can be performed using keywords similar to the conventions used in PHREEQC, eliminating the need to prepare the database file for every new system. Another modification was the application of DQMOM, which offers several advantages in terms of speed, robustness, and adaptability over our previously implemented method (QMOM). Subtle technicalities in the implementation of DQMOM to obtain a reliable and quickly converging solution method were explained to allow replication/extension of the current work by other researchers. We also provide fully commented MATLAB codes implementing the PBE simulation workflow in the accompanying Supporting Information.

Upon developing an improved computational framework, three different global uncertainty/sensitivity analysis (UA/SA) methods were applied to understand the behavior of the model in response to uncertainty in various model parameters. For several simulation outputs, either we demonstrated the consistency of the results from different SA measures or explained the reason behind the inadequacy of the applied method. In the latter case, for instance, we presented particle edge length as an output whose highly skewed distribution limited the applicability of variance-based indices. Having a comprehensive picture of the uncertainty propagation through the model, we simplified the variability space of the model parameters and employed UA/SA as a tool for the theory-driven design of nanoparticle formation processes. Here, we showed that—within simplifying assumptions such as the constancy of C-S-H composition across the considered experimental conditions—one could take advantage of UA/SA to decide on the optimal synthesis conditions for a target property in the final product. The procedure was demonstrated using the specific surface area of particles as the output, revealing the critical role of reagent addition rate and intercrystallite cohesion energy in obtaining the desirable outcome. These results warrant similar future studies using mechanistic kinetic models with carefully calibrated parameters that are only dependent on the system specifications. Such models allow for UA/SA over a wide range of operating conditions, which subsequently provide an invaluable basis for the rational design of production units. In the particular case of C-S-H system examined here, future experimental work relating the effect of different factors—most importantly, mixing/hydrodynamics as well as additives such as polymers and/or salts—on the properties of the product could lead to the development of more predictive models, which in turn guide the synthesis more decisively. In a broader sense, the current study provides a basis for future efforts where meticulously developed computational models guide the practical implementation in the laboratory.

## Acknowledgements

This work was supported by the Energy and Environment Research Division (ENE), Paul Scherrer Institute, Switzerland.

The authors thank Dr. Sasa Bjelic (PSI) for providing additional computational resources, Prof. Christian Ludwig (PSI and EPFL) and Dr. Debora Foppiano (formerly at PSI) for valuable discussions, Dr. Laurin Wissmeier (CSD INGENIEURE AG) and Prof. D. Andrew Barry (EPFL) for providing their MATLAB codes for interfacing with IPhreeqc, and Prof. Karen Scrivener (EPFL) for continued support.

## Associated content

**Supporting Information Available:** Additional details on computational procedures, supplementary uncertainty/sensitivity analysis results, and fully commented MATLAB codes implementing the population balance modelling workflow. This material is available free of charge via the Internet at http://pubs.acs.org.

# Supporting Information

# Global Uncertainty-Sensitivity Analysis on Mechanistic Kinetic Models: From Model Assessment to Theory-Driven Design of Nanoparticles

M. Reza. Andalibi[*,†,‡], Paul Bowen[‡], Agnese Carino[†], Andrea Testino[†]

[†] Paul Scherrer Institute (PSI), ENE-LBK-CPM, Villigen-PSI, Switzerland.

[‡] École polytechnique fédérale de Lausanne (EPFL), STI-IMX-LMC, Lausanne, Switzerland.

[*] Corresponding author: reza.andalibi@psi.ch

## 1.     Population Balance Equation and Its Solution Using DQMOM

Consider a population of crystallites whose size distribution is evolving over time. The number density function (NDF; $n$) describing this distribution is a function of time ($t$) and crystallite characteristic size ($L$), and can be approximated by a discretized distribution as

$$n(L,t) = \sum_{\alpha=1}^{N} w_\alpha(t)\delta\bigl(L - L_\alpha(t)\bigr) \qquad (1)$$

where $\alpha$ denotes the discretization nodes with weights $w_\alpha$ at sizes (or abscissas) $L_\alpha$, $N$ is the overall number of the nodes, and $\delta$ is the Dirac delta function [1,2]. For a homogeneous system (namely, one with uniformity across the physical space), the temporal evolution in NDF can be expressed using the so-called population balance equation (PBE)

$$\frac{\partial n}{\partial t} = S(L) \qquad (2)$$

where $S$ is a source term embracing all the solid formation/transformation processes such as nucleation, growth, and aggregation, inflows and outflows of crystallites, and possible changes in the volume of reaction liquor. In the kinetic modelling of precipitation processes, the PBE is solved along with differential equations written for mass balances (*viz.*, the conservation of elements inside the reactor).

Substituting Eq. (1) in Eq. (2) (the time-dependences are dropped for simplicity)

$$\sum_{\alpha=1}^{N} \frac{dw_\alpha}{dt}\delta(L - L_\alpha) - \sum_{\alpha=1}^{N} w_\alpha \frac{dL_\alpha}{dt}\delta'(L - L_\alpha) = S(L) \qquad (3)$$

where $\delta'(L - L_\alpha)$ is the first derivative of the generalized delta function [3]. Now, defining the weighted abscissas

$$\varsigma_\alpha = w_\alpha L_\alpha \qquad (4)$$

we will have the following ODE upon substitution in Eq. (3)

$$\sum_{\alpha=1}^{N} \frac{dw_\alpha}{dt} \delta(L - L_\alpha) - \sum_{\alpha=1}^{N} \left(\frac{d\varsigma_\alpha}{dt} - L_\alpha \frac{dw_\alpha}{dt}\right) \delta'(L - L_\alpha) = S(L) \tag{5}$$

Defining

$$a_\alpha \equiv \frac{\partial w_\alpha}{\partial t}$$
$$b_\alpha \equiv \frac{\partial \varsigma_\alpha}{\partial t} \tag{6}$$

followed by substitution in Eq. (5), gives

$$\sum_{\alpha=1}^{N} a_\alpha [\delta(L - L_\alpha) + L_\alpha \delta'(L - L_\alpha)] - \sum_{\alpha=1}^{N} b_\alpha \delta'(L - L_\alpha) = S(L) \tag{7}$$

Applying the moment transformation defined as

$$m_k(t) \equiv \int_0^\infty L^k \, n(t, L) dL \cong \sum_{\alpha=1}^{N} w_\alpha(t) L_\alpha^k \tag{8}$$

and knowing [3]

$$\int_0^\infty L^k \delta(L - L_\alpha) \, dL = L_\alpha^k \tag{9}$$

$$\int_0^\infty L^k \delta'(L - L_\alpha) dL = -k L_\alpha^{k-1} \tag{10}$$

we transform Eq. (7) by multiplying both sides by $L_\alpha^k$ and integrating over $[0, \infty)$ yielding

$$(1 - k) \sum_{\alpha=1}^{N} L_\alpha^k a_\alpha + k \sum_{\alpha=1}^{N} L_\alpha^{k-1} b_\alpha = \int_0^\infty L^k S(L) dL \tag{11}$$

Now, defining the moment source term $\bar{S}_k$

$$\bar{S}_k \equiv \int_0^\infty L^k S(L) dL \tag{12}$$

Eq. (11) can be recast into a matrix form as (boldface symbols are vectors and matrices)

$$\boldsymbol{A\alpha} = [\boldsymbol{A_1}, \boldsymbol{A_2}] \begin{bmatrix} \boldsymbol{a} \\ \boldsymbol{b} \end{bmatrix} = \boldsymbol{d} \tag{13}$$

where

$$\boldsymbol{A_1} = \begin{bmatrix} 1 & \cdots & 1 \\ 0 & \cdots & 0 \\ -L_1^2 & \cdots & -L_N^2 \\ \vdots & \vdots & \vdots \\ (1-k)L_1^k & \cdots & (1-k)L_N^k \\ \vdots & \vdots & \vdots \\ 2(1-N)L_1^{2N-1} & \cdots & 2(1-N)L_N^{2N-1} \end{bmatrix}_{2N \times N} \tag{14}$$

$$A_2 = \begin{bmatrix} 0 & \cdots & 0 \\ 1 & \cdots & 1 \\ 2L_1 & \cdots & 2L_N \\ \vdots & \vdots & \vdots \\ kL_1^{k-1} & \cdots & kL_N^{k-1} \\ \vdots & \vdots & \vdots \\ (2N-1)L_1^{2N-2} & \cdots & (2N-1)L_N^{2N-2} \end{bmatrix}_{2N \times N} \quad (15)$$

$$\boldsymbol{\alpha} = [a_1, a_2, \ldots, a_N, b_1, b_2, \ldots, b_N]^T = \begin{bmatrix} \boldsymbol{a} \\ \boldsymbol{b} \end{bmatrix}_{2N \times 1} \quad (16)$$

and for a well-mixed homogeneous system (namely, no dependence on physical space)

$$\boldsymbol{d} = [\bar{S}_0, \bar{S}_1, \ldots \bar{S}_k, \ldots, \bar{S}_{2N-1}]^T \quad (17)$$

Note that since $A$ is not dependent on $w_\alpha$ it can be defined even for nodes with zero weight.

At this juncture, let us elaborate on some technical issues associated with DQMOM and their resolution. For nanoparticle formation and transformation systems, matrix $A$ can be extremely ill-conditioned, complicating the accurate solution of the linear system in Eq. (13) [4]. The first reason for this ill-conditioning can readily be understood by looking at the rows of matrix $A$ which are composed of abscissas raised to nonnegative integers. Working with SI units and with crystallite sizes in the order of nanometers ($10^{-9}$ m), the rows of matrix $A$ will assume very different scales spanning several orders of magnitude (*e.g.*, 44 and 35 orders of magnitude variability in $A_1$ and $A_2$, respectively for $N$=3). To alleviate this problem, we can follow the abscissas in nanometers. Thus, defining $L_{\alpha,nm} \equiv L_\alpha \times 10^{-9}$ and substitution in Eq. (11)

$$(1-k) \sum_{\alpha=1}^{N} L_{\alpha,nm}^k a_\alpha \times 10^{-9k} + k \sum_{\alpha=1}^{N} L_{\alpha,nm}^{k-1} b_\alpha \times 10^{-9(k-1)} = \int_0^\infty L^k S(L) dL \quad (18)$$

Moreover, substitution in Eq. (6) gives $b_{\alpha,nm}$ as

$$b_\alpha \equiv \frac{d(w_\alpha L_\alpha)}{dt} = \frac{d(w_\alpha L_{\alpha,nm} \times 10^{-9})}{dt} = 10^{-9} \times b_{\alpha,nm} \quad (19)$$

Inserting this unit adjusted variable into Eq. (18) yields

$$(1-k) \sum_{\alpha=1}^{N} L_{\alpha,nm}^k a_\alpha + k \sum_{\alpha=1}^{N} L_{\alpha,nm}^{k-1} b_{\alpha,nm} = 10^{9k} \int_0^\infty L^k S(L) dL \quad (20)$$

which in matrix form reads

$$\boldsymbol{A}_{nm} \boldsymbol{\alpha}_{nm} = [\boldsymbol{A}_{1,nm}, \boldsymbol{A}_{2,nm}] \begin{bmatrix} \boldsymbol{a} \\ \boldsymbol{b}_{nm} \end{bmatrix} = \boldsymbol{d}_{nm} \quad (21)$$

with

$$A_{1,nm} = \begin{bmatrix} 1 & \cdots & 1 \\ 0 & \cdots & 0 \\ -L_{1,nm}^2 & \cdots & -L_{N,nm}^2 \\ \vdots & \vdots & \vdots \\ (1-k)L_{1,nm}^k & \cdots & (1-k)L_{N,nm}^k \\ \vdots & \vdots & \vdots \\ 2(1-N)L_{1,nm}^{2N-1} & \cdots & 2(1-N)L_{N,nm}^{2N-1} \end{bmatrix}_{2N \times N} \quad (22)$$

$$A_{2,nm} = \begin{bmatrix} 0 & \cdots & 0 \\ 1 & \cdots & 1 \\ 2L_{1,nm} & \cdots & 2L_{N,nm} \\ \vdots & \vdots & \vdots \\ kL_{1,nm}^{k-1} & \cdots & kL_{N,nm}^{k-1} \\ \vdots & \vdots & \vdots \\ (2N-1)L_{1,nm}^{2N-2} & \cdots & (2N-1)L_{N,nm}^{2N-2} \end{bmatrix}_{2N \times N} \quad (23)$$

$$\boldsymbol{\alpha}_{nm} = [a_1, a_2, \ldots, a_N, b_{1,nm}, b_{2,nm}, \ldots, b_{N,nm}]^T = \begin{bmatrix} \boldsymbol{a} \\ \boldsymbol{b}_{nm} \end{bmatrix}_{2N \times 1} \quad (24)$$

$$\boldsymbol{d}_{nm} = [\bar{S}_0, \bar{S}_1 \times 10^9, \ldots \bar{S}_k \times 10^{9k}, \ldots, \bar{S}_{2N-1} \times 10^{9(2N-1)}]^T \quad (25)$$

In our experience, the scaling procedure explained earlier is an inevitable step in the application of DQMOM to the process of nanoparticle formation. Yet another reduction in the condition number of matrix $A_{nm}$ can be readily achieved by preconditioning which makes the convergence of the iterative solution more robust and faster [5]. This is particularly important when dealing with particulate processes that give rise to sharp changes in the particle phase space (*e.g.*, nucleation or aggregation) [4]. Here, we applied a left preconditioning using a diagonal matrix with main diagonal elements

$$P_{ii} = \frac{\bar{L}_{nm}^{i-1} + \bar{L}_{nm}^{i-2}}{2}; i = 1, \ldots, 2N \quad (26)$$

where $\bar{L}_{nm}$ is the average of the abscissas in nm

$$\bar{L}_{nm} = \frac{1}{N} \sum_{\alpha=1}^{N} L_{\alpha,nm} \quad (27)$$

Therefore, instead of solving the system of linear equations in Eq. (21) we solve

$$\boldsymbol{P}^{-1} \boldsymbol{A}_{nm} \boldsymbol{\alpha}_{nm} = \boldsymbol{P}^{-1} \boldsymbol{d}_{nm} \quad (28)$$

With $\boldsymbol{P}$ being diagonal, its inverse $\boldsymbol{P}^{-1}$ can trivially be obtained *via* inverting the main diagonal elements [3].

After these considerations, solving Eq. (28) at each time step of integrating the set of ordinary differential equations (ODE set composed of PBE + mass balances) yields $\boldsymbol{a}$ and $\boldsymbol{b}_{nm}$. For the ODE solvers to work efficiently, the dependent variables have to be properly scaled [6]. Here, having the weights in the range of $10^{20}$ crystallites.m$^{-3}$ (or higher), both $a_\alpha = \frac{\partial w_\alpha}{\partial t}$ and $b_{\alpha,nm} = \frac{\partial (w_\alpha L_{\alpha,nm})}{\partial t}$ can be extremely large, especially during the burst of nucleation (nucleation rates are in excess of $10^{16}$ crystallites.m$^{-3}$.s$^{-1}$ for typical model parameters [7]). To bring these values to the order of unity, we solved the ODE set for $\log_{10}$ of the weights and weighted abscissas

$$\frac{d(\log_{10} w_\alpha)}{dt} = \frac{dw_\alpha}{dt} \times \frac{1}{w_\alpha \ln 10} \tag{29}$$

$$\frac{d(\log_{10} \varsigma_{\alpha,nm})}{dt} = \frac{d\varsigma_{\alpha,nm}}{\partial dt} \times \frac{1}{\varsigma_{\alpha,nm} \ln 10} \tag{30}$$

where $\varsigma_{\alpha,nm} \equiv w_\alpha L_{\alpha,nm}$.

Now, let us describe the constituents of the source term $\bar{S}_k$. For molecular growth (which could be size-dependent [7]) in a homogenous system we have [1]

$$S_G(L) = -\frac{\partial}{\partial L}[n(L,t)G(L)] \tag{31}$$

Therefore,

$$\bar{S}_k^G = \int_0^\infty L^k S_G(L) dL = k \int_0^\infty L^{k-1} G(L) \times n(L,t) dL \tag{32}$$

which was obtained using integration by parts [3]. Now, using Eq. (1)

$$\bar{S}_k^G = k \sum_{\alpha=1}^N w_\alpha L_\alpha^{k-1} G(L_\alpha) \tag{33}$$

In fact, this is the N-point quadrature approximation of the growth source term for moment order k [1,7]. In matrix format and for abscissas in nanometers

$$\boldsymbol{d_{nm}^G} = \begin{bmatrix} 0 \\ \sum_{\alpha=1}^N w_\alpha G(L_\alpha) \times 10^9 \\ \vdots \\ k \sum_{\alpha=1}^N w_\alpha L_\alpha^{k-1} G(L_\alpha) \times 10^{9k} \\ \vdots \\ (2N-1) \sum_{\alpha=1}^N w_\alpha L_\alpha^{(2N-2)} G(L_\alpha) \times 10^{9(2N-1)} \end{bmatrix} =$$

$$\boldsymbol{diag}(1, 10^9, \dots, 10^{9k}, \dots, 10^{9(2N-1)}) \times \boldsymbol{A_2} \times \boldsymbol{diag}(w_1, w_2, \dots, w_N) \times \begin{bmatrix} G(L_1) \\ G(L_2) \\ \vdots \\ G(L_N) \end{bmatrix} \tag{34}$$

where $\boldsymbol{diag}()$ denotes a diagonal matrix.

For nucleation, the source term is [1]

$$S_N(L) = J_{hom}\delta(L - L_{hom}^*) + J_{sec}\delta(L - L_{sec}^*) \tag{35}$$

where $J_{hom}$ and $J_{sec}$ are the rates of primary homogeneous and true secondary nucleation processes (crystallites.m$^{-3}$.s$^{-1}$) while $L_{hom}^*$ and $L_{sec}^*$ are the respective critical nucleus sizes (in m). Therefore, the corresponding moment source term ($\bar{S}_k^N$) is

$$\bar{S}_k^N = \int_0^\infty L^k S_N(L) dL = L_{hom}^{*\,k} J_{hom} + L_{sec}^{*\,k} J_{sec} \tag{36}$$

Again, in matrix form

$$\boldsymbol{d}_{nm}^N = \boldsymbol{diag}(1, 10^9, \ldots, 10^{9k}, \ldots, 10^{9(2N-1)}) \times \begin{bmatrix} 1 & 1 \\ L_{hom}^* & L_{sec}^* \\ \vdots & \vdots \\ L_{hom}^{*\,k} & L_{sec}^{*\,k} \\ \vdots & \vdots \\ L_{hom}^{*\,2N-1} & L_{sec}^{*\,2N-1} \end{bmatrix} \times \begin{bmatrix} J_{hom} \\ J_{sec} \end{bmatrix} \tag{37}$$

Finally, the source term for changes in the system volume (for flow systems) reads

$$\bar{S}_k^{Volume} = -m_k \frac{d(\ln V)}{dt} \tag{38}$$

where $V$ is the volume of the reaction suspension. Again, in matrix form

$$\boldsymbol{d}_{nm}^{Volume} = -\frac{d(\ln V)}{dt} \boldsymbol{diag}(1, 10^9, \ldots, 10^{9k}, \ldots, 10^{9(2N-1)}) \times \begin{bmatrix} m_0 \\ m_1 \\ \vdots \\ m_{2N-1} \end{bmatrix} \tag{39}$$

Another complication arises from the fact that PBE only tracks the evolution of crystallites. Therefore, an additional differential equation is required to account for the time variation of particle number concentration ($n_p$; particles.m$^{-3}$) [7,8]. Again, with $n_p$ adopting very large values we solve the ODE for $\log_{10}$ of $n_p$

$$\frac{d(\log_{10} n_p)}{dt} = \frac{dn_p}{dt} \times \frac{1}{n_p \ln 10} \tag{40}$$

$$\frac{dn_p}{dt} = -n_p \frac{d(\ln V)}{dt} + J_{hom} \tag{41}$$

Besides the PBE set and the ODE for $n_p$, we have to solve for mass balances over elemental abundances ($n_e$) inside the solution. Since in the current precipitation system the molar amounts are in mmol range the derivatives are defined in terms of mmol.s$^{-1}$ to bring them closer to the order of unity. The rate of precipitate formation therefore provides the coupling between kinetics and thermodynamic speciation calculation at each time step of integrating the ODE set [7,9,10].

## 2. Additional Details on the Implementation of PBE Simulation Framework

Throughout our PBE simulations, we generally used 10$^{-12}$ and 10$^{-5}$, respectively, for the relative and absolute tolerances input to the MATLAB's ODE solver (ode15s). Moreover, to avoid convergence problems with PHREEQC we used a maximum step size of 3 (modified using the keyword data block KNOBS, keyword -step_size; default value is 100). This fine-tunes the variation in the activities of master species in a single iteration and make the solution more robust [11]. Despite these considerations, particularly in the case of very sharp fronts (as with, for instance, extremely fast nucleation at very high $Q$ or very low $\gamma$ accompanied by high $\sigma/\gamma$), ODE solver may converge to less accurate results or may not converge at all. This can readily be checked by comparing the precipitation yield calculated from the third moment against that estimated from Ca elemental balance. The latter value is invariably more reliable as the calculation of moments in moment-based methods is always associated with some error which generally increases with the moment order and depends on the specific problem [2,12]. In our experience, with the default tolerances the discrepancy between the two yields is generally low and within a few percent. Nevertheless, an improvement can be achieved by reducing the absolute tolerance to $10^{-6}$. This, however, comes at the expense of additional computational time. Therefore, in all our simulations we first used a larger tolerance ($10^{-5}$) and redid the simulation with smaller tolerance only for cases with yields different by more than 1%.

Another important detail about the implementation of moment-based methods concerns the initial conditions for the solution of population balance equations. When there are no particles in the reactor at t = 0, the initial conditions for the weights and weighted abscissas would be zero [1]. Thus, in the first time step the abscissas are not defined, the matrix $A$ is not invertible, and hence, the terms $a$ and $b$ cannot be calculated. A viable way to work around this issue is to introduce a negligible amount of tiny fictitious seeds in the reactor to start the integration and maintain stability during the computations [1,7,10]. In the current work, $N \times 10^4$ m$^{-3}$ crystallites/particles with sizes $(1, 2, …, N) \times 10^{-10}$ m were used to seed the PBE simulations.

One last implementation detail is related to the simulations performed at temperatures other than the room temperature. In the absence of thermodynamic data on the temperature dependence of C-S-H dissolution reaction (in particular for Ca:Si = 2), we adopted an approximation protocol recommended for the so-called isocoulombic reactions [13,14]. Such reactions have the same number of ions on each side of the reaction, or at least have the same total charge on both sides. This simplifies the temperature dependence of the reaction because of variations in $C_p^\circ$ for different species canceling out [13,14]. Additionally, Gu et al. [14] noticed that not only $\Delta C_p^\circ$ and $\Delta S^\circ$ terms are small for isocoulombic reactions, but also they are usually of opposite signs canceling each other out. This leads to the so-called one-term extrapolation approach which states that for well-balanced aqueous reactions $\Delta_r G^\circ$ does not depend on temperature. This way, only the Gibbs energy change of reaction at one temperature (usually room temperature) suffices to estimate the value at other temperature [14]. The method also applies to isocoulombic reactions involving condensed phases with non-reference state solids (mostly minerals) giving less accurate estimates compared to when the solid is in its reference state [14]. For reactions that are not inherently isocoulombic, combining with other reactions (whose equilibrium constants are known as a function of temperature) is a necessary step before using the one-term extrapolation method [13]. For this purpose, oftentimes the ionization of water is used. For our C-S-H solid, the dissolution reaction reads

$$(CaO)_{1.0}(SiO_2)_{0.5}(H_2O)_{1.5}(s) \leftrightharpoons Ca(OH)^+ + 0.5HSiO_3^- + 0.5OH^- + 0.5H_2O;$$
$$\log_{10} K_1^{25°C} = -7.22 \qquad (42)$$

Adding that to

$$0.5H^+ + 0.5OH^- \leftrightharpoons 0.5H_2O; \log_{10} K_2^{25°C} = 7.00 \tag{43}$$

and

$$0.5H^+ + 0.5HSiO_3^- \leftrightharpoons 0.5SiO_2(aq) + 0.5H_2O; \log_{10} K_3^{25°C} = 4.90 \tag{44}$$

gives

$$(CaO)_{1.0}(SiO_2)_{0.5}(H_2O)_{1.5}(s) + H^+ \leftrightharpoons Ca(OH)^+ + 0.5SiO_2(aq) + 1.5H_2O;$$
$$\log_{10} K_4^{25°C} = 4.68 \tag{45}$$

which is an isocoulombic reaction and the temperature dependence for its equilibrium constant can be approximated using the one-term correlation of Gu et al [14]. Therefore,

$$\log_{10} K_4^T = \frac{298.15}{T} \log_{10} K_4^{25°C} = \frac{1395.342}{T} \tag{46}$$

Now, knowing the temperature dependences of reactions (43), (44), and (45) (the first two from CEMDATA18 [15]), we can "un-combine" the equilibrium constant of reaction (1) as a function of temperature. To provide the data into PHREEQC, we regressed a line to $\log_{10} K_1^T$ vs. $1/T$ data over a temperature range 10-50°C. The enthalpy change of the reaction (42) can then be back calculated from either the slope or the intercept (according to the van't Hoff equation [11]), giving a value of 14.85 kJ/mol.

```matlab
%% Experimental system: C-S-H precipitation in semi-batch reaction mixing Ca(NO3)2 with Na2SiO3
T = 25; % Experimental temperature ('C)

% List of species, and other thermodynamic information-----------------------------------------------%
% List of elements including those not initially in the system but added later, for example to keep pH constant
% (e.g., if NaOH is not present initially but is added during the raeaction to regulate pH)
elements = {'Ca','Si','Na','N','Ntg'};
% List of aqueous species (those in which we're interested) compatible with database used in PHREEQC
aqSpecies = ...
    {'OH-','H+','Ca+2','Ca(OH)+','CaSiO3','Ca(HSiO3)+',...
    'HSiO3-','SiO3-2','SiO2','Si4O10-4','Na+','NaOH','NO3-','Ntg'};
% List of gaseous species with known partial pressures
gasSpecies = {'Ntg(g)'};
% A flag with ones in the first row denoting the elements that have to be equilibrated forcing
% desired partial pressure of the gas with the corresponding index in the second row
elements_gasEquilFlag = [0,0,0,0,1; 0,0,0,0,1];
% List of solid species; at the moment only the FIRST one can be precipitated in the kinetic code
solidSpecies = {'CSH','Portlandite'};
% Index of solid phases not already in the thermodynamic database and/or defined by the user;
% the precipitating solid has to be the first solid
solidDefInd = 1;

% Inputs for defining the added solid phase
% Dissolution reaction
dissRxn_precSolid = {'(CaO)1.0(SiO2)0.5(H2O)1.5 = Ca(OH)+ + 0.5HSiO3- + 0.5OH- + 0.5H2O'};
% Molar volume
molarVol_precSolid = 49.20e-6; % m3/mol
% log10 of equilibrium constant for the dissolution reaction above
logK_precSolid = -7.21958906895829;
% Enthalpy of dissolution reaction above (J/mol)
delta_h_precSolid = 14850;

% Experimental conditions (concentrations, volumes, flow rate, etc)----------------------------------%
% Initial volumes and concentration are reported at room T
% Elements initially in the reactor, rxt, (mol/kg H2O or approximately mol/L for dilute solutions);
% for elements only related to a gas phase (like Ntg) put zero because the amount is calculated by fixing
% the partial pressure
element_rxtConc = [0.02,0,0,0.04,0];
% Elements added to the reactor with the inflow (mol/kgw molality in the stream)
element_inflowConc = [0,9e-3,0.117,0,0];
% log10 of partial pressures for the gas phase species (vector for multiple gases)
gasSI = 0;
% Initial volume of solution inside the reactor (m3)
V0_rxt = 0.2e-3;
% Total volume of the inflow solution to be added (m3)
V0_inflow = 0.222e-3;
% Flow rate of the inflow stream (m3/s)
Q_inflow = 0.5e-6/60;
% Leave empty if pH is not fixed; give a pH value if fixed
set_pH = [];
% The name of compound used to keep the pH constant; all the elements of this species have to be present in "elements"
fixed_pH_Compound = 'NaOH';
% Atom symbol of the ion of pH-fixing compound the amount of which has to be adjusted (by charge balance) to attain
% fixed pH; has to be present in "elements"
fixed_pH_Elem = 'Na';
```

Figure S 1. A MATLAB script containing typical input for thermodynamic calculations in PHREEQC.

# 3. Supplementary UA/SA Results with Model Parameters as Input Factors

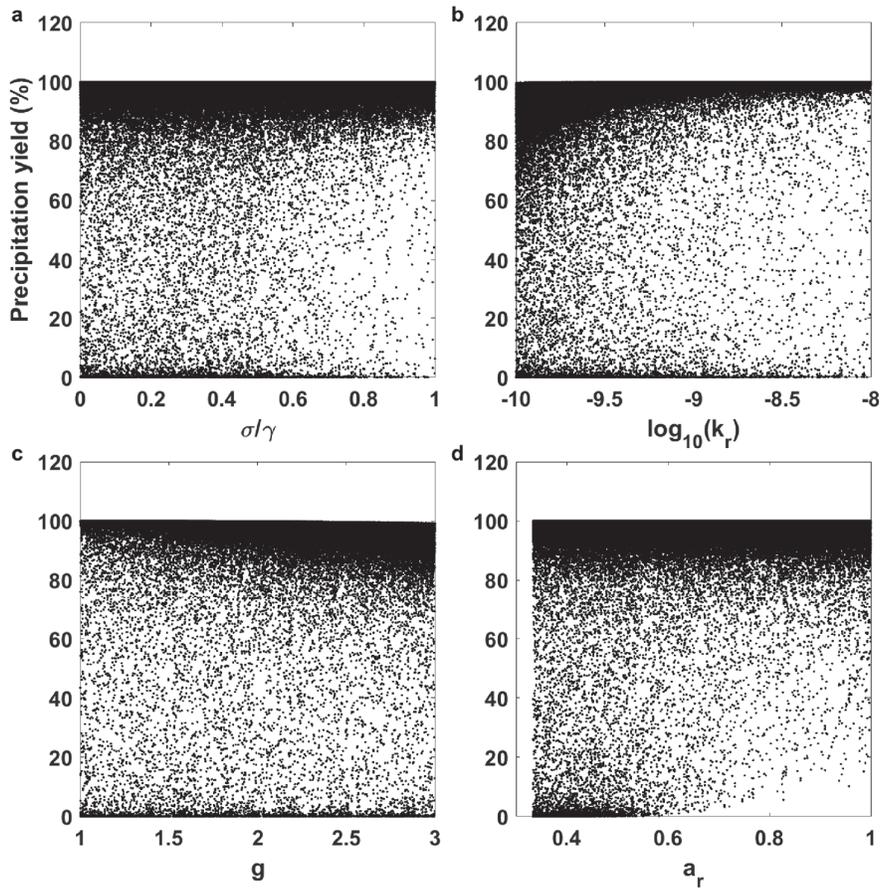

Figure S 2. Scatter plots of precipitation yield with respect to equilibrium composition *vs.* $\sigma/\gamma$, $\log_{10} k_r$, $g$, and $a_r$.

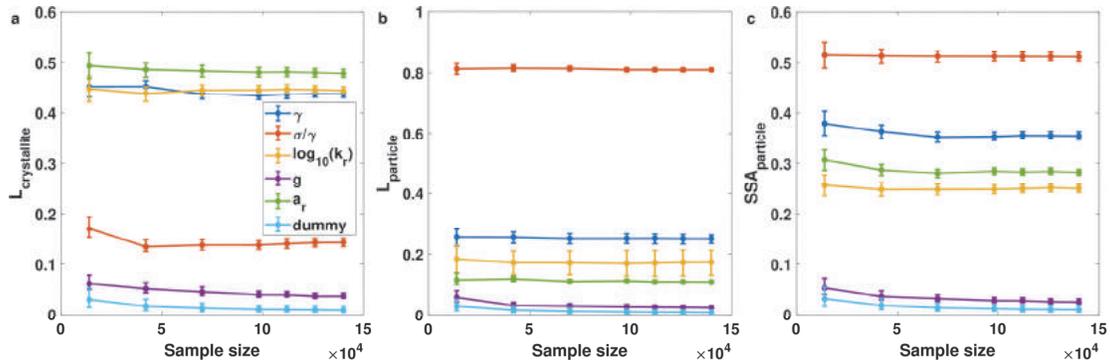

Figure S 3. The convergence of PAWN sensitivity indices (maximum KS statistic) *vs.* sample size for the three different model outputs crystallite thickness (a), particle edge length (b), and particle surface area (c).

Figure S 3 summarizes the convergence analysis results for the maximum KS statistic as the sensitivity index. As we can see, the convergence is achieved already at sample sizes above 50,000 and the 95% confidence intervals are narrow, allowing for a straightforward assessment of relative importance exhibited by various inputs. Overall, the relatively fast convergence of different indices, narrow confidence intervals,

and the consistency of the results with scatter plots testify to the suitability of PAWN for the current UA/SA problem.

Figure S 4 summarizes the convergence behavior for the average of absolute EEs (Figure S 4(a-c)) and their C.V. values (Figure S 4(d-f)). For the mean values convergence is obtained at sample sizes above 60,000 in all case and the confidence intervals are quite narrow (Figure S 4 (a-c)). Achieving convergence for the C.V. values, on the contrary, poses a significant challenge and the confidence intervals are relatively wide even at a very large sample size of 120,000 (Figure S 4 (d-f)). Fixing $g$ to a nominal value 2 and estimating the EET indices over a sample of size 195,000 resolves this issue (Figure S 5). With this larger sample, the obtained measures and their C.V. do not change appreciably but the confidence intervals generally become smaller (compare Figure 6 to Figure S 5). In particular, with $\bar{L}_c$ as the output, the confidence interval of C.V. for $\log_{10} k_r$ separates from that of other parameters assuming the lowest value among others (Figure S 5(a)). Similarly, with $SSA_p$ as the output, the C.V. for $a_r$ separates from that of $\gamma$ and $\log_{10} k_r$ (Figure S 5(c)).

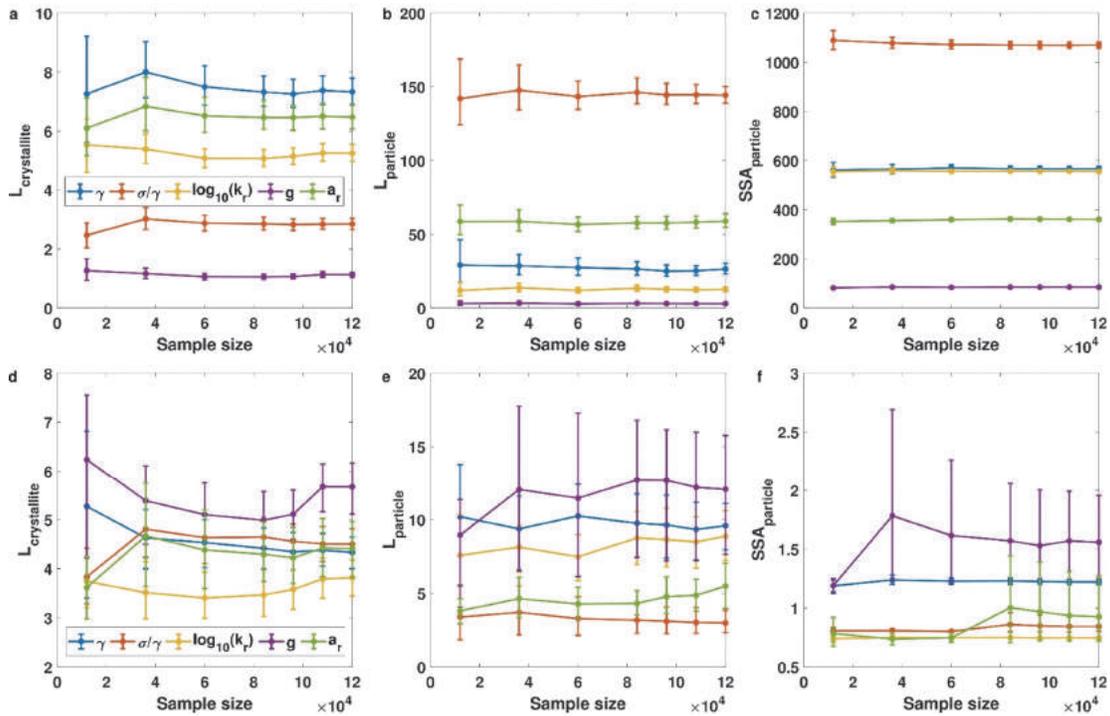

Figure S 4. (a-c) convergence plots for the average of Elementary Effects as a function of sample size; (d-f) convergence plots for coefficients of variation *versus* sample size (all the 95% confidence intervals are estimated by bootstrapping).

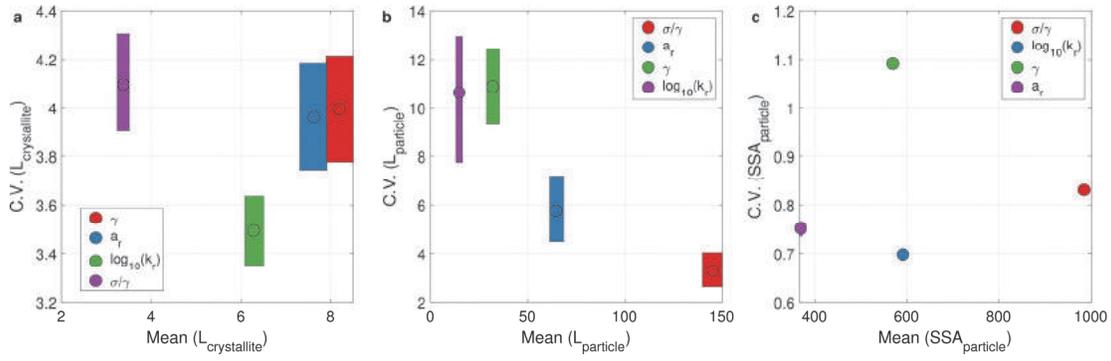

Figure S 5. Sensitivity indices obtained using EET with crystallite thickness (a), particle edge length (b), and particle surface area (c) as the outputs (sample size of 195,000; $g$=2).

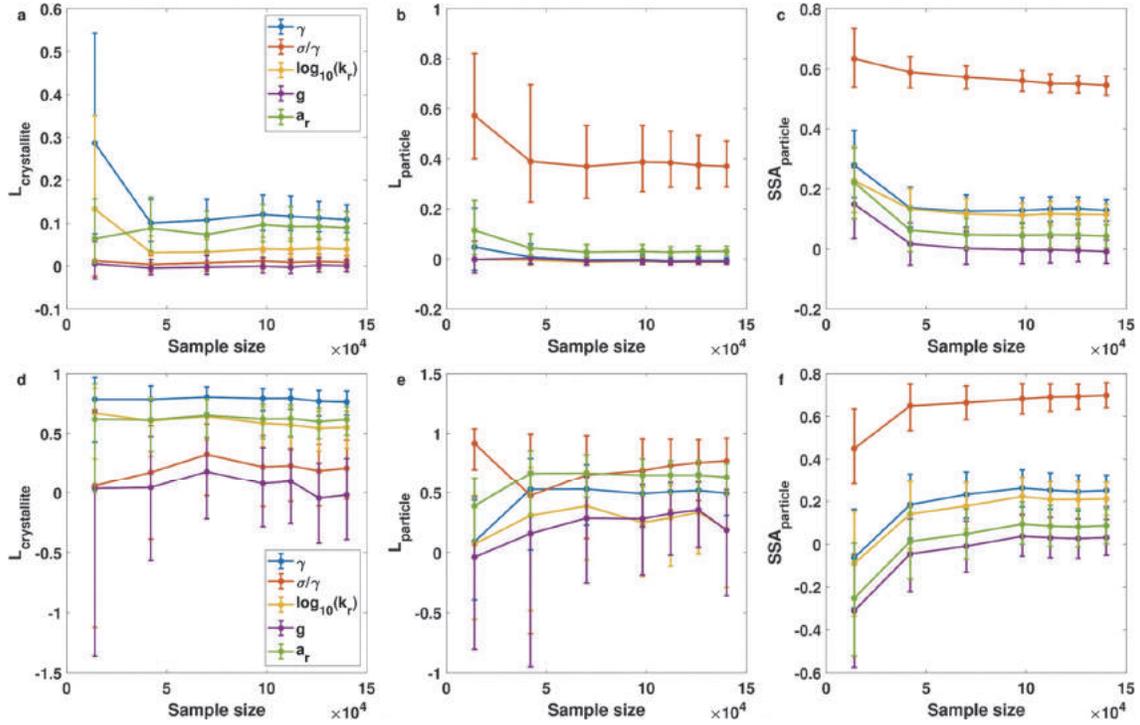

Figure S 6. Convergence behavior of variance-based main effects (a-c) and total effects (d-f) with untransformed outputs ($\bar{L}_c$, $\bar{L}_p$, and $SSA_p$) as a function of input sample size, with error bars representing the 95% confidence intervals.

Figure S 6 summarizes the convergence of variance-based indices. From Figure S 6(a-c) we notice a relatively fast convergence for the main effects. Specifically, concerning $\bar{L}_p$ and $SSA_p$ both having one parameter with much larger main effect ($\sigma/\gamma$) the convergence is reached at sample sizes in excess of 50,000 (Figure S 6(b,c)). Still with $\bar{L}_c$ the convergence is fast despite having parameters with similar main effects (Figure S 6(a)). On the contrary, convergence of the total effects introduces a significant challenge in the case of $\bar{L}_c$ and $\bar{L}_p$ (Figure S 6(d,e)) given their highly skewed distributions as we discussed earlier (Figure 3(a,b)). This mainly manifests in wide and significantly overlapping confidence intervals while the average values from bootstrapping already stabilize for larger sample sizes (Figure S 6(d,e)). This is more evident when $g$ is fixed to a nominal value of 2 and the analysis is extended to even larger sample sizes (Figure S

9(d,e)). For $SSA_p$ having a well-behaved probability distribution, the convergence of total effects is not a problem as we can see in Figure S 6(f) (and Figure S 9(f)).

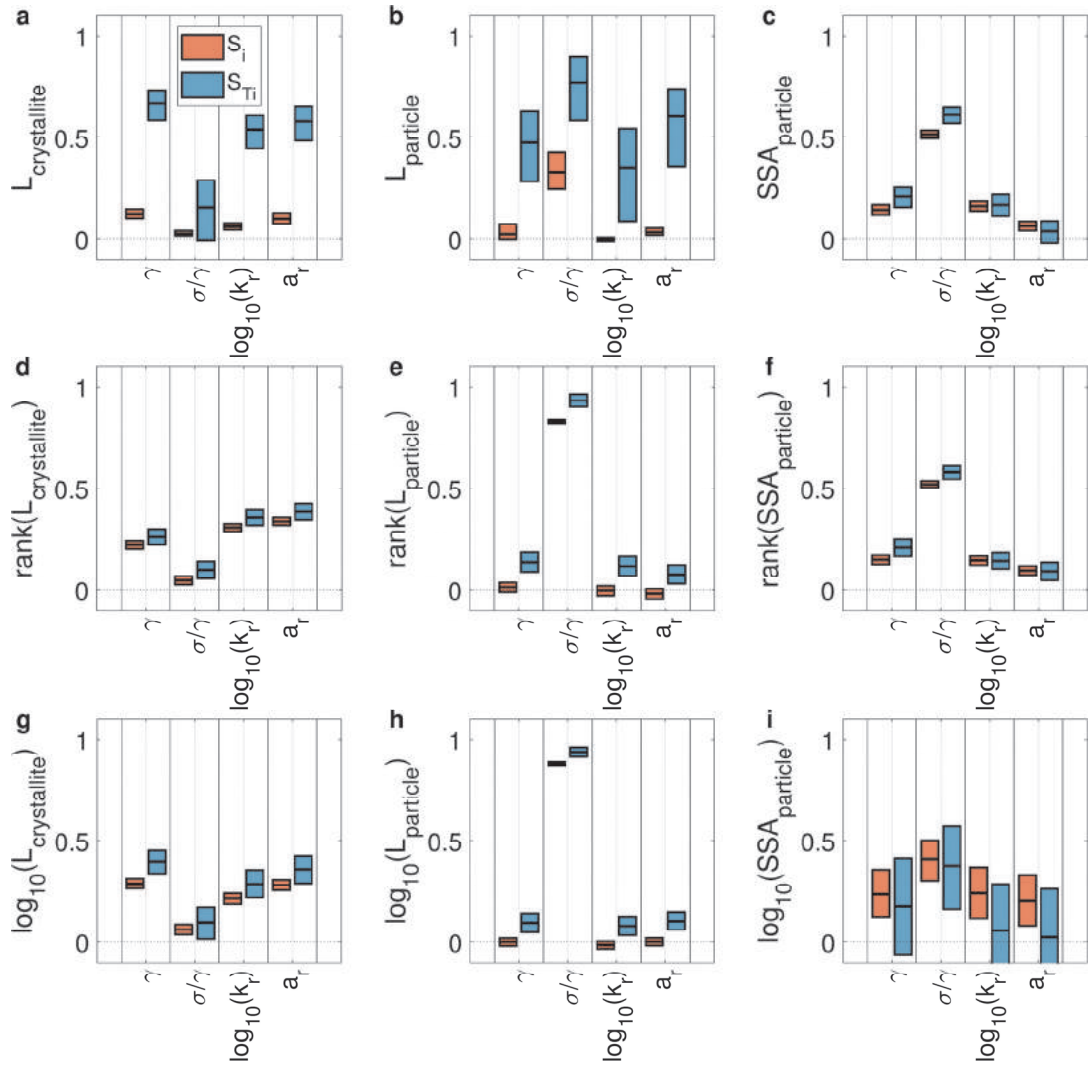

Figure S 7. Variance-based sensitivity indices with fixed kinetic order of growth ($g$ = 2), for $\bar{L}_c$, $\bar{L}_p$, and $SSA_p$ (a-c), their rank transformations (d-f), and log$_{10}$ transformations (g-i) as the outputs, with a sample size 288000 (base sample size $N$ = 48000).

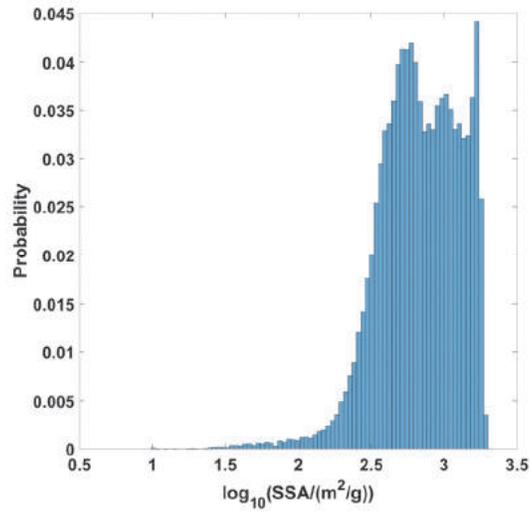

Figure S 8. Probability-normalized histograms of $\log_{10} SSA_p$.

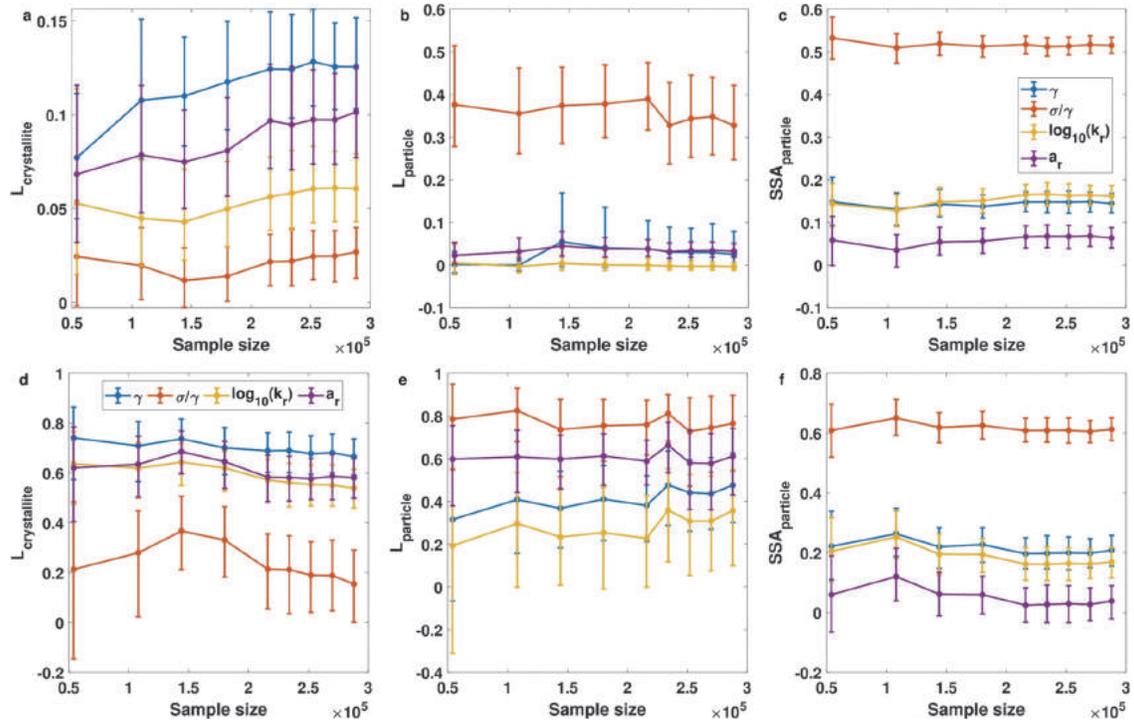

Figure S 9. Convergence behavior of variance-based main effects (a-c) and total effects (d-f) with untransformed outputs ($\bar{L}_c$, $\bar{L}_p$, and $SSA_p$) as a function of input sample size, with error bars representing the 95% confidence intervals; the parameter $g$ is fixed to a nominal value of 2.

# 4. Supplementary UA/SA Results with Selected Model Parameters and Experimental Conditions as Input Factors

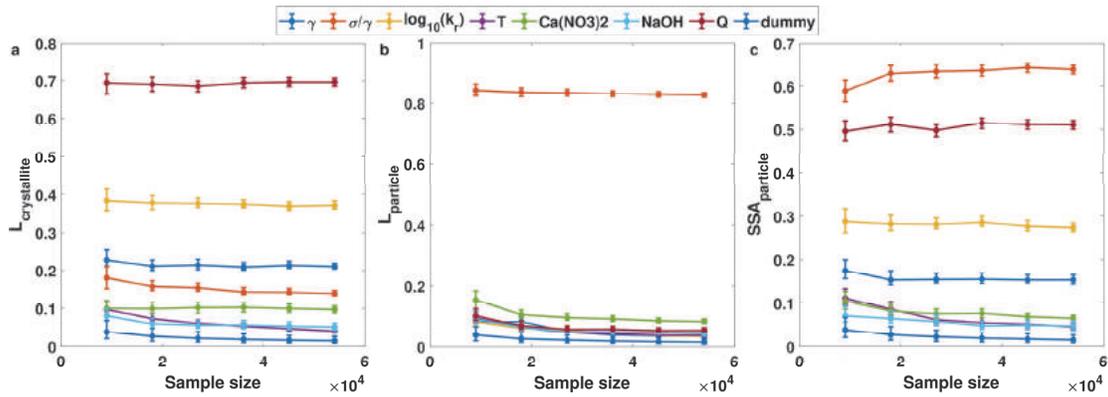

Figure S 10. The convergence of PAWN sensitivity indices (maximum KS statistic) *vs.* sample size for the three different model outputs crystallite thickness (a), particle edge length (b), and particle surface area (c) in the UA/SA with selected model parameters and experimental conditions as the input factors.

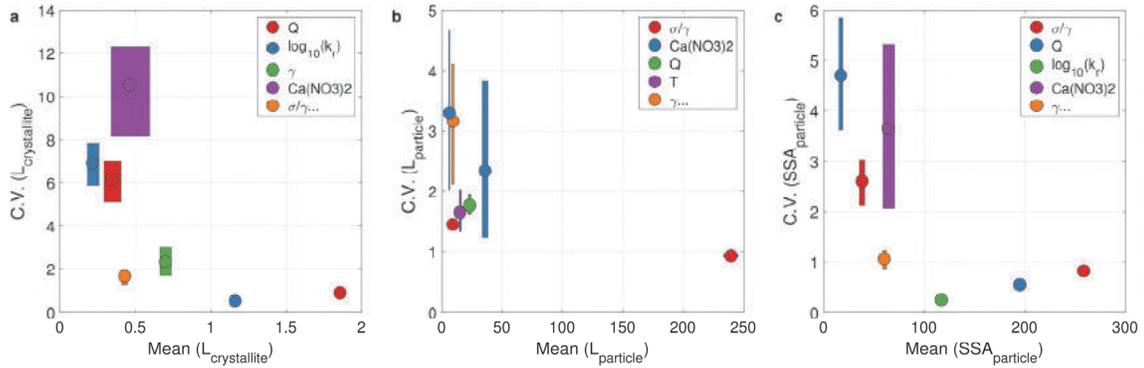

Figure S 11. Results of SA with selected model parameters and experimental conditions as the input factors. Sensitivity indices obtained using EET with crystallite thickness (a), particle edge length (b), and particle surface area (c) as the output (sample size of 48,000). The results are presented as the mean of Elementary Effects plotted against their coefficients of variation.

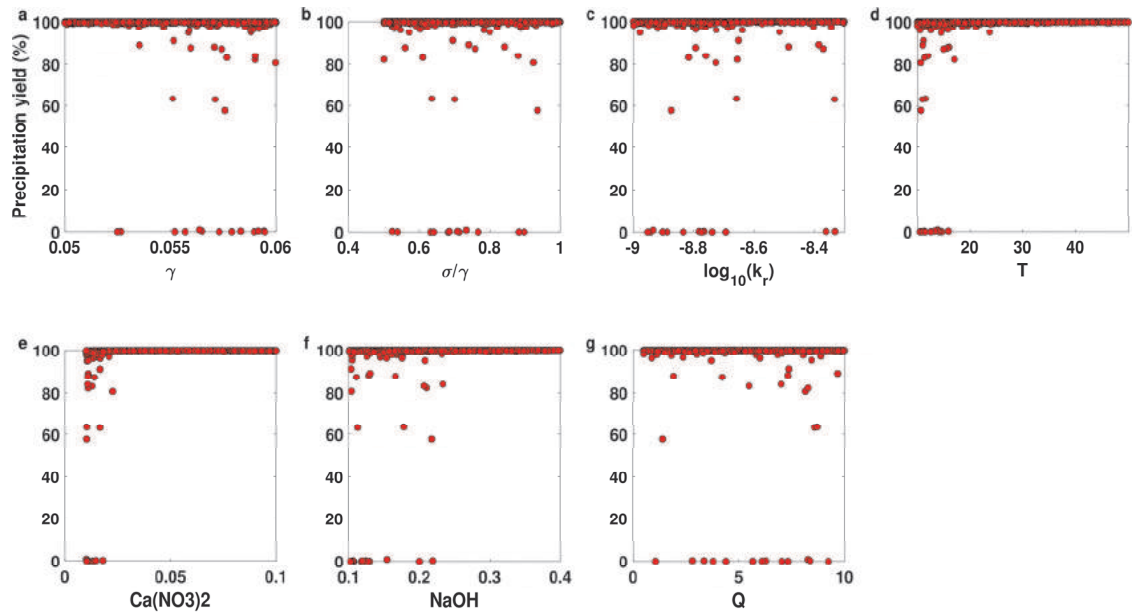

Figure S 12. Scatter plots of precipitation yield *vs.* different factors in UA with selected model parameters and experimental conditions as the uncertain inputs.

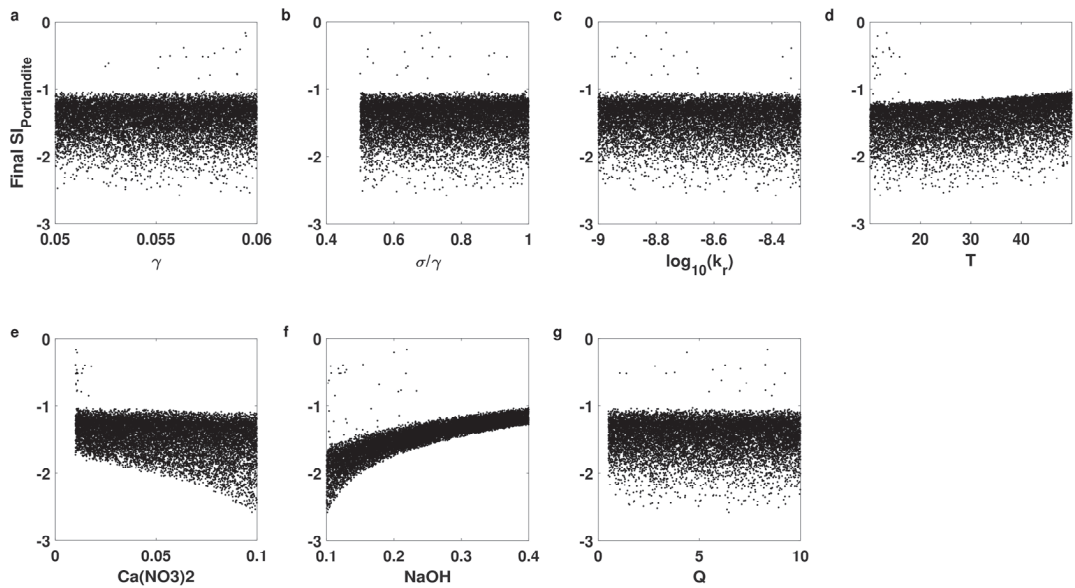

Figure S 13. Dependence of final saturation index with respect to portlandite on different factors in UA with selected model parameters and experimental conditions as the uncertain inputs.

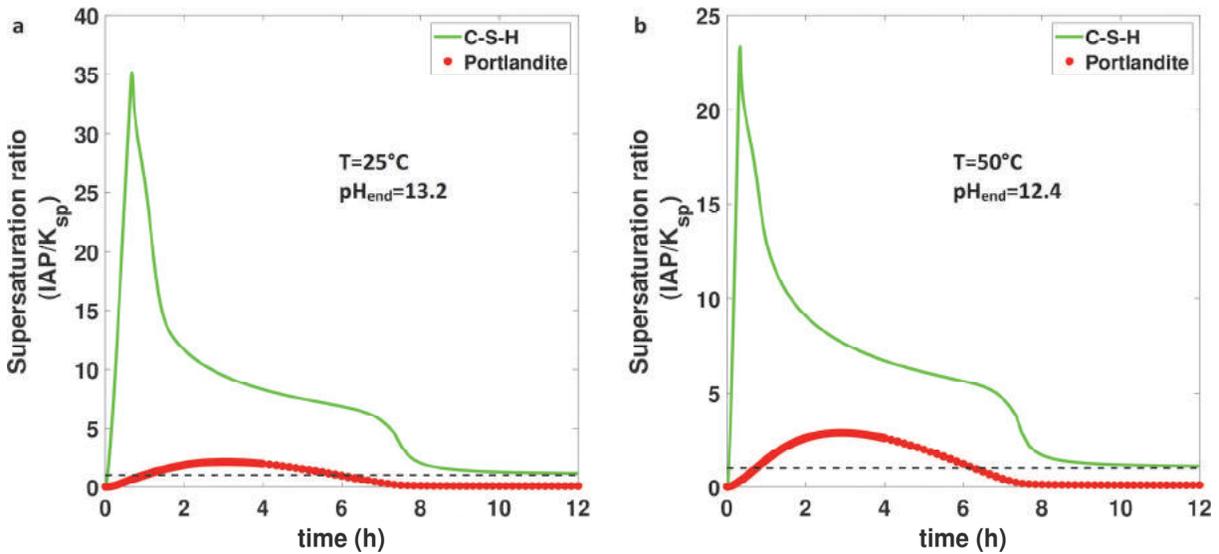

Figure S 14. Temporal supersaturation ratio with respect to C-S-H and portlandite during a precipitation process at room temperature (a) and at 50°C (b) with a NaOH concentration of 0.4 mol/kg water. All the kinetic parameters and the rest of experimental conditions are the same as in Ref. [7] (experiment at $Q$=0.5 mL.min$^{-1}$) and the simulations allowed only for the precipitation of C-S-H (the horizontal dashed line in black denotes equilibrium condition with supersaturation ratio of unity).

From Figure S 13(d,f) we see that the chance of precipitating portlandite along with C-S-H increases both with the NaOH concentration in the inflow stream and the experimental temperature. This can readily be understood by looking at the precipitation reaction of portlandite ($Ca^{2+} + 2OH^- \leftrightharpoons Ca(OH)_2$) which is an endothermic process and is promoted at higher $OH^-$ activities [15]. A less strong influence can also be discerned with decreasing the Ca(NO3)$_2$ concentration (Figure S 13(e)) although this parameter has to be kept high enough to achieve acceptable precipitation yields (Figure S 12(e)). PBE simulations at NaOH concentration 0.4 mol/kg water (4 times the value in our experiments in Ref. [7]) at room temperature and 50°C show these effects, giving rise to periods of supersaturation ratio in excess of unity for portlandite (Figure S 14). Fortunately, none of these parameters directly influence the C-S-H particle SSA and therefore, it should be possible to produce high surface area C-S-H product without coprecipitating portlandite.